\documentclass[prl,twocolumn,floatfix,superscriptaddress,amsmath,citeautoscript,aps,longbibliography]{revtex4-2}
\pdfoutput=1
\usepackage[T1]{fontenc}\usepackage[latin1]{inputenc}
\usepackage{dcolumn,bm,graphicx,color,booktabs,microtype,afterpage} \graphicspath{{Figures/}}
\renewcommand{\figurename}{Fig.}
\renewcommand{\tablename}{Table}
\makeatletter\renewcommand{\fnum@figure}[1]{\figurename~\thefigure.}\makeatother
\makeatletter\renewcommand{\fnum@table}[1]{\tablename~\thetable.}\makeatother
\newcount\hh \newcount\mm
\hh=\time \divide\hh by 60
\mm=\hh \multiply\mm by 60 \mm=-\mm
\advance\mm by \time
\def\now{\number\hh:\ifnum\mm<10{}0\fi\number\mm}
\usepackage[colorlinks,plainpages=false,linkcolor=blue,urlcolor=blue,citecolor=blue,pdfpagemode=UseNone,pdfstartview=FitBH]{hyperref}
\usepackage{xcolor}	
	\usepackage{amsmath,amsfonts,amsbsy,amssymb,amsthm}
	\usepackage{comment}
	\usepackage{tikz}
	\usepackage{esint}
	\usepackage{amsopn}
	\usepackage{tipa}
	\usepackage{braket}
	\usepackage{bm}
	\usepackage{hyperref}
	\usepackage{epsfig,float}
	\usepackage{times}
	\usepackage{verbatim}   
	\usepackage{color}      
	\hypersetup{
		    unicode=false,          
		    pdftoolbar=false,        
		    pdfmenubar=true,        
		    pdffitwindow=false,     
		    pdfstartview={FitH},    
		    pdfkeywords={MBL} {AGP} {Mass-deformed SYK}, 
		    pdfnewwindow=true,      
		    colorlinks=true,       
		    linkcolor=red,          
		    citecolor=blue,        
		    filecolor=magenta,      
		    urlcolor=blue           
		}
	\usepackage{xcolor}
	\usepackage{graphicx}
	\usepackage{dcolumn}
	\usepackage{bm}
	\usepackage{epsfig}
	\usepackage{psfrag}
	\usepackage{orcidlink}
	
	\newcommand{\reffig}[1]{\mbox{Fig.~\ref{#1}}}
	\newcommand{\refeq}[1]{\mbox{Eq.~(\ref{#1})}}

	\newcommand{\be}{\begin{equation}}
	\newcommand{\ee}{\end{equation}}
	\newcommand{\bal}{\begin{align}}
	\newcommand{\eal}{\end{align}}
	\def\bea{\begin{eqnarray}}
	\def\eea{\end{eqnarray}}
	\renewcommand{\Re}{\mathrm{Re}}

	\newcommand{\pcsadd}{Center for Theoretical Physics of Complex Systems, Institute for Basic Science (IBS), Daejeon 34126, Republic of Korea}
		\newcommand{\ustadd}{Basic Science Program, Korea University of Science and Technology (UST), Daejeon 34113, Republic of Korea}

	\newcommand{\apctpadd}{Asia Pacific Center for Theoretical Physics, Pohang, Gyeongbuk, 37673, Republic of Korea}

	\newcommand{\vietnamadd}{Institute for Interdisciplinary Research in Science and Education, ICISE, Quy Nhon, Vietnam}

\begin{document}
			
	\title{Nonrelativistic versus relativistic quantum scars in billiard systems}

		\author{Barbara Dietz,\orcidlink{0000-0002-8251-6531}}
		    \email{Corresponding Author: bdietzp@gmail.com}
		    \affiliation{\pcsadd}
		    \affiliation{\ustadd}

		\author{Dung Xuan Nguyen,\orcidlink{0000-0002-8595-0528}}
		    \email{dungmuop@gmail.com}
		    \affiliation{\pcsadd}
		    \affiliation{\ustadd}
		    \affiliation{\vietnamadd}

		\author{Tilen \v{C}ade\v{z},\orcidlink{0000-0002-5343-4086}}
		    \email{tilen.cadez@apctp.org}
		    \affiliation{\pcsadd}
		    \affiliation{\apctpadd}

		\date{\today}

		\begin{abstract}
			We study the features of scarred eigenstates of relativistic neutrino billiards (NBs), graphene billiards (GBs) and Haldane graphene billiards (HGBs) and recapitulate those for nonrelativistic quantum billiards (NRQBs) with the shapes of a full- and quarter-stadium billiard. Here, we restrict for the GBs and HGBs to the region of linear dispersion around the Fermi energy, where they are effectively described by the same Dirac equation for massless spin-1/2 particles as NBs. Scarred wave functions of the nonrelativistic billiards and spinor functions of the relativistic ones are localized along the same types of periodic orbits, the most prominent ones being bouncing-ball orbits. The objective is to demonstrate that the properties of the scarred eigenstates observed in the full- and quarter-stadium GB \emph{do not comply} with those of relativistic quantum systems. For this we apply the semiclassical approach associated with such non-generic contributions, which was developed for the spectral density of NRQBs and NBs. It provides semiclassical trace formulas in terms of the periodic orbits associated with a scarred wave function and a procedure to extract such contributions from the eigenvalue spectra. Furthermore, we analyze momentum distributions and Husimi functions of such scarred states and employ them to classify scarred wave functions according to the periodic orbits along which they are localized. We show that for the GB the semiclassical approach, the spectral properties, the symmetry properties and generally properties of the wave functions all comply with those of the NRQB, whereas for the HGB they agree well with those of the NB. Thus, even though around the Fermi energy GBs are described by the relativistic Dirac equation the quantum scars, or generally, the quantum scarred eigenstates observed in GBs do not exhibit those of relativistic ones.  
		\end{abstract}
		\bigskip
		\maketitle

\section{Introduction\label{Intro}}
	We investigate quantum scars, or more generally, scarred wave function in neutrino billiards (NBs), graphene billiards (GBs) and Haldane-graphene billiards (HGBs) with the shapes of full- and quarter-stadium billiards and compare them with those of nonrelativistic quantum billiards (NRQBs)~\cite{Heller1984,Bogomolny1988,Berry1989,Sieber1993,Tomsovic1993,Bohigas1993,Baecker1997,Kaplan1998,Kaplan1999,Bies2001,Selinummi2024,Graf2026} and other systems~\cite{Rahkonen2017,Rahkonen2019,Rahkonen2025}. These were found experimentally for the first time in flat microwave resonators~\cite{Sridhar1991,Stein1992} in the range of microwave frequencies where the Helmholtz equation governing these systems is mathematically identical to the Schr\"odinger equation. The classical Bunimovich stadium billiard~\cite{Sinai1970,Bunimovich1979,Berry1981,Berry1983} exhibits full chaos and periodic orbits are typically unstable and isolated and cover the whole available phase space. Yet, an exception is a non-generic, continuous family of neutral periodic orbits of measure zero, named bouncing-ball orbits (BBOs), which bounce back and forth between the two straight-line segments. These yield an additional contribution, derived analytically in Ref.~\cite{Sieber1993}, to Gutzwiller's trace formula~\cite{Gutzwiller1971,Gutzwiller1990}, which provides a semiclassical approximation for the spectral density of quantum systems with fully chaotic classical dynamics. In Ref.~\cite{Sieber1993} semiclassical trace formulas were also derived for other periodic orbits that lead to a scarring of wave functions of the stadium billiard. Especially the presence of BBOs leads to deviations of the long-range correlations in the eigenvalue spectra from random-matrix theory (RMT) predictions~\cite{Berry1977b,Casati1980,Bohigas1984} for typical classically chaotic systems. Namely, according to the Bohigas, Giannoni and Schmit conjecture~\cite{Bohigas1984} the spectral properties of generic quantum systems with chaotic classical dynamics are well described by those of random matrices from the Gaussian ensembles of corresponding universality class~\cite{Mehta1990,LesHouches1989,Guhr1998,Haake2018}, which is the Gaussian orthogonal (GOE) and unitary (GUE) ensemble for preserved and violated time-reversal invariance, respectively. For generic quantum systems with an integrable classical dynamics~\cite{Robnik1998} the spectral properties are the same as for Poissonian random numbers~\cite{Berry1977a}. The semiclassical theory which shows that for generic systems the level statistics are universal and follow the predictions of RMT was developed by Berry in Ref.~\cite{Berry1985,Sieber1993,Heusler2007} based on periodic orbit theory~\cite{Gutzwiller1971,Gutzwiller1990} for the spectral density. Here, a crucial assumption is that all the periodic orbits of the system are isolated and unstable, which is not fulfilled in the stadium billiard due to the presence of the continuous non-generic set of neutral BBOs in the stadium billiard. Indeed, it serves as a paradigm model for studies of aspects of quantum chaos and the effect of scarred wave functions on the spectral properties~\cite{Graef1992,Sieber1993}. Besides providing a semiclassical expression for the level density originating from the BBOs in Ref.~\cite{Sieber1993}, a procedure was developed for the extraction of the non-generic effects of the BBOs on the spectral properties. It was extended to relativistic NBs in Ref.~\cite{Dietz2020,Dietz2022} and employed to extract contributions from scarred eigenstates in NBs with semicircular shape, shapes with a threefold symmetry and constant-width NBs in Ref.~\cite{Zhang2021,Yupei2022,Dietz2022}. Neutrino billiards were proposed by Berry and Mondragon in Ref.~\cite{Berry1987}. Their spinor eigenstates are solutions of the relativistic Weyl equation~\cite{Weyl1929} -- generally referred to as two-dimensional Dirac equation in context of NBs -- for a massless spin-1/2 particle subject to the BC that there is no outgoing flow.

\begin{figure}
	\centering
	\begin{tabular}{@{}c@{}}
	\begin{tabular}{@{}c@{}}
	\includegraphics[width=.48\linewidth]{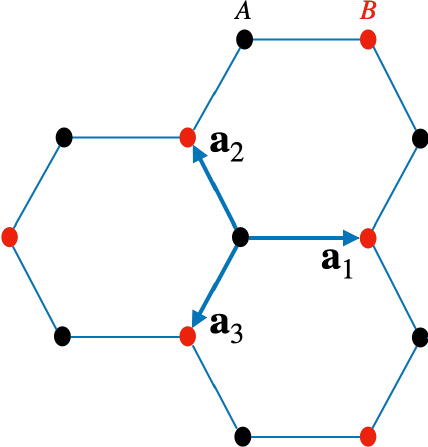} \\[\abovecaptionskip]
	\end{tabular}
	\begin{tabular}{@{}c@{}}
	\includegraphics[width=.49\linewidth]{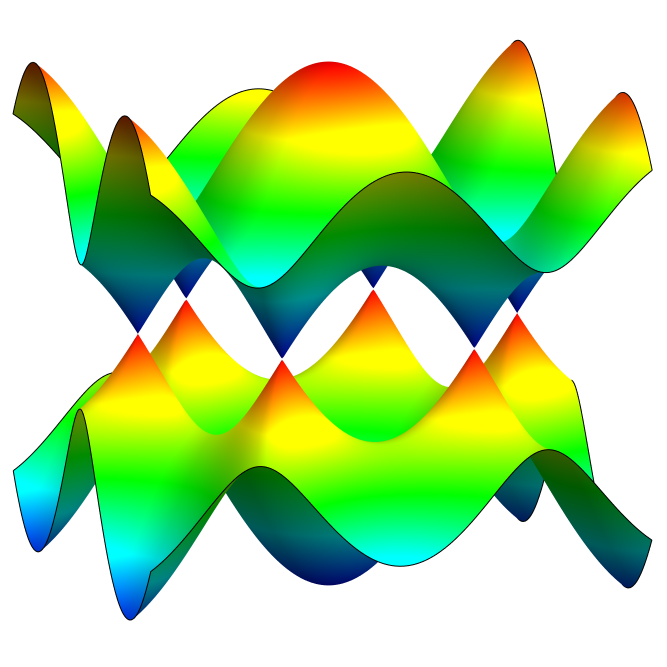}
	\end{tabular}
	\end{tabular}
	\caption{Left: The honeycomb structure of graphene. Right: The conduction and valence bands of graphene resulting from tight-binding model calculations. They touch each other conically at the corners of the first Brillouin zone.}\label{BandStr_GB}
	\end{figure}

We report on studies of the properties of the eigenstates of GBs~\cite{DiVincenzo1984,Novoselov2004,Geim2007,Avouris2007,Miao2007,Ponomarenko2008,Beenakker2008,Zhang2008,Neto2009,Abergel2010,Zandbergen2010} and HGBs~\cite{Nguyen2024} with the shapes of a full-stadium and a quarter-stadium. We construct the GBs by cutting out of a honeycomb lattice a sheet with the desired shape and compute their eigenstates based on a tight-binding model~\cite{Dietz2015}. The valence and conduction bands of graphene touch each other conically at the Fermi energy at the corners of the hexagonal Brillouin zone as illustrated in~\reffig{BandStr_GB}, implicating there a linear dispersion relation. In that region the energy excitations of graphene are well described by the Dirac equation for massless spin-1/2 particles~\cite{Wallace1947,Novoselov2004,Geim2007,Beenakker2008,Neto2009}. The touch points $K$ and $K^\prime$ associated with the two triangular lattices are referred to as 'Dirac points' in the following. Boundary conditions (BCs) to be imposed in that region on the spinor components of a GB are given in Refs.~\cite{Akhmerov2007,Akhmerov2008,Wurm2011}. From these we select Dirichlet BCs on the next-nearest outer sites along the boundary~\cite{Dietz2015}. The conical shape originates from the honeycomb lattice structure~\cite{Slonczewski1958}, which is formed by two interpenetrating triangular sublattices and the presence of two Dirac cones originates from time-reversal symmetry, inversion symmetry, and the discrete rotational symmetry $C_3$ of the honeycomb-lattice structure~\cite{Beenakker2008,Bernevigbook2013}. The analogy to the wave equation of relativistic systems led to numerous numerical~\cite{Wurm2009,Huang2010,Rycerz2012,Rycerz2013,Dietz2017a,Zhang2021,Zhang2023c} and experimental 'artificial-graphene' realizations~\cite{Polini2013,Parimi2004,Joannopoulos2008,Bittner2010,Kuhl2010,Singha2011,Nadvornik2012,Gomes2012,Tarruell2012,Uehlinger2013,Rechtsmann2013,Rechtsmann2013a,Khanikaev2013,Wang2014a,Shi2015,Bellec2013,Dietz2015,Dietz2016,Zhang2023}.
		
	We restrict to the conical valley regions around the Dirac points, where GBs are effectively described by the same Dirac Hamiltonian as NBs and exhibit relativistic phenomena, like pseudodiffusive transport, the quantum Hall effect, Zitterbewegung, Klein tunneling and edge states~\cite{DiVincenzo1984,Novoselov2004,Geim2007,Avouris2007,Miao2007,Ponomarenko2008,Beenakker2008,Zhang2008,Neto2009,Abergel2010,Zandbergen2010}. Accordingly, it was originally expected that due to this analogy, in the region of linear dispersion their spectral properties are similar to those of NBs of corresponding shape, that is, that they exhibit GUE statistics if their shape has no mirror symmetry and corresponds to that of a chaotic billiard. This expectation was confirmed in experiments with graphene quantum dots~\cite{Guettinger2008,Guettinger2010,Ponomarenko2008}, whereas numerical studies~\cite{Libisch2009,Wurm2009} and experimental studies of GBs with superconducting microwave Dirac billiards~\cite{Dietz2015,Dietz2016,Zhang2023} revealed that they conform with those of NRQBs of corresponding shape, that is, with GOE statistics. In Refs.~\cite{Rycerz2012,Rycerz2013} these discrepancies were attributed to the BCs, and in Ref.~\cite{Wurm2009} it was shown, that the reason is back scattering at the boundary, which leads to a mixing of valley states around the $K$ and $K^\prime$ points~\cite{Wurm2009}. Note, that time-reversal invariance is violated at each Dirac cone, where the electronic excitations are effectively described by the same relativistic Dirac equation~\cite{Beenakker2008} as massless NBs. However, time-reversal invariance is restored due to the occurrence of two independent Dirac cones that are mapped onto each other when applying the associated time-reversal operator. 
	\begin{figure}
	\centering
	\begin{tabular}{@{}c@{}}
	\begin{tabular}{@{}c@{}}
	\includegraphics[width=.35\linewidth]{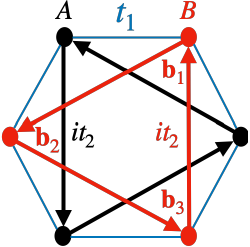} \\[\abovecaptionskip]
	\end{tabular}
	\begin{tabular}{@{}c@{}}
	\includegraphics[width=.49\linewidth]{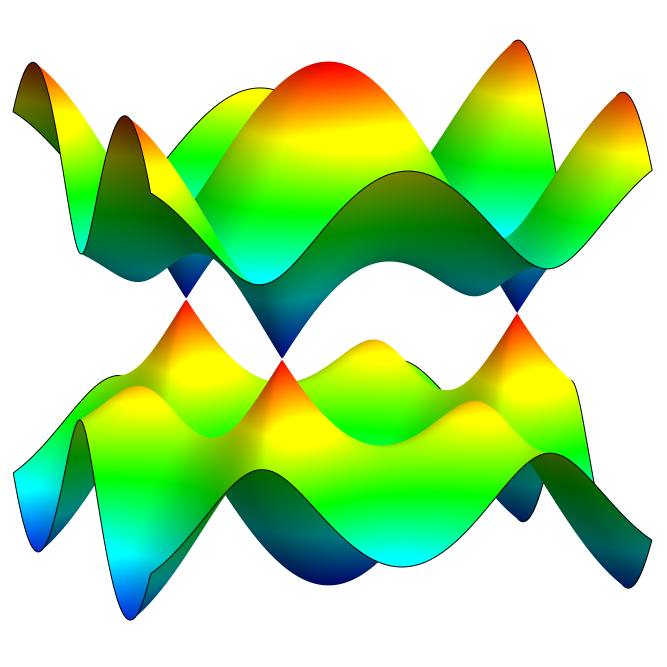} 
	\end{tabular}
	\end{tabular}
		\caption{Same as~\reffig{BandStr_GB} for graphene subject to the Haldane-model onsite potential $M=0.3$ and next-nearest neighbor tunneling at the critical point $t_2=M/(3\sqrt{3})$.}\label{BandStr_HB}
	\end{figure}

This interpretation was confirmed in Ref.~\cite{Nguyen2024} where we introduced a gap at the $K^\prime$ point so that within the energy range of the gap the eigenstates are confined to the valley region around the $K$ point; cf.~\reffig{BandStr_HB}. This is achieved by applying the Haldane model~\cite{Haldane1988,KaneMele2005} to the honeycomb lattice of a GB. Specifically, by appropriately tuning the Semenoff mass $M$, generated by opposite onsite potentials on the $A$ and $B$ sublattices and thus breaking inversion symmetry, and the Haldane mass arising from purely imaginary next-nearest-neighbor hopping, the resulting band structure exhibits a single Dirac cone at either the $K$ or ithe $K'$ point, while the other valley remains gapped.  Due to the purely imaginary next-to-nearest neighbor tunneling term time-reversal invariance is explicitely broken in a HGB in the region of linear dispersion relation. On the contrary, close to the band edges, the eigenvalues of the HGB are close to those of the corresponding GB, and after appropriate scaling also to those of the NRQB of corresponding shape. We demonstrated in Ref.~\cite{Nguyen2024} that for such critical values within the energy range of the gap, the spectral properties of a HGB are similar to those of the corresponding NB~\cite{Berry1987}. 

It is general belief that the quantum scars observed in stadium GBs are relativistic~\cite{Huang2009,Xu2013,Ge2024,Dietz2025}, even though it is established by now, that spectral properties are the same as those of the corresponding NRQB. We demonstrate in this article, that scarred states of full- and quarter-stadium shaped GBs agree with those of the NRQB, whereas those of the HGB coincide with those of the corresponding relativistic NB. For this we study for all four billiard types spectral properties and properties of scarred wave functions in terms of the momentum distributions and for the NRQBs and NBs also Husimi functions. Furthermore we apply the semiclassical approach for the contributions of non-generic modes to the fluctuating part of the spectral density, and find for all measures that features of scarred eigenstates and generally eigenstates of GBs and HGBs are well captured by those of the corresponding NRQB and NB, respectively. We even find, that the wavenumbers of BBOs in the GB are equal to those of the NRQB and in the HGB they agree well with those of the NB. 

\section{Nonrelativistic and Relativistic Quantum Billiards~\label{NRQBNB}}
\subsection{Brief review of the salient differences of NRQBs and NBs}
	The domain $\Omega$ of the stadium billiard, composed of a rectangular part with side lengths $2L$ and two semicircular parts with radius $r_0$, is defined as $\boldsymbol{r}=[X(\xi,\eta),Y(\xi,\eta)]$, or in the complex plane as $w(\xi,\eta)=X(\xi,\eta)+iY(\xi,\eta)$. Here, we choose polar coordinates $(\xi=r,\eta=\phi)$, $r\leq r_0, \phi\in [0.2\pi)$ in the semicircular parts, $X=r\cos\phi\pm L, Y=r\sin\phi$, and cartesian coordinates $(\xi=x,\eta=y), -L\leq x\leq L,-r_0\leq y \leq r_0$ in the rectangular part, $X=x, Y=y$. Accordingly, the boundary $\partial\Omega$ is defined by setting $r=r_0$ and $y=\pm r_0$, respectively. Here we chose for the ratio of $L$ and $r_0$ the golden mean, $\frac{L}{r_0}=\frac{1+\sqrt{5}}{2}$. 

	The eigenstates of the NRQBs are obtained by solving the Schr\"odinger equation of a free particle with Dirichlet BCs along $\partial\Omega$,
	\bea
	\hat H_S\psi(\xi,\eta)&&=-\Delta_{(\xi,\eta)}\psi(\xi,\eta)=k^2\psi(\xi,\eta),\,
	\label{Schr}\\
	\psi(\xi,\eta)\vert_{\partial\Omega}&&=0,\nonumber
	\eea
	where $k$ denotes the wavenumber associated with the energy $E=k^2$. The eigenstates of NBs~\cite{Berry1987} are obtained by solving the Weyl equation~\cite{Weyl1929} for a massless, non-interacting spin-1/2 particle, commonly referred to as Dirac equation in that context,
	\be
	\hat H_D\boldsymbol{\psi}(\boldsymbol{r})=c\boldsymbol{\hat\sigma}\cdot\boldsymbol{\hat p}\boldsymbol{\psi}(\boldsymbol{r})
	=E\boldsymbol{\psi}(\boldsymbol{r}),\, \boldsymbol{\psi}(\boldsymbol{r})=
	\begin{pmatrix}
	\psi_1(\boldsymbol{r}) \\ \psi_2(\boldsymbol{r})
	\end{pmatrix},
	\label{DE}
	\ee
	with $\boldsymbol{\hat p}=-i\hbar\boldsymbol{\nabla}$ denoting the momentum of the particle, $\hat H_D$ the Dirac Hamiltonian and  $E=\hbar ck$, and imposing the condition, that the outgoing flux, that is, the normal component of the local current,
	\be
	\boldsymbol{n}\cdot\boldsymbol{u}(\boldsymbol{r})=c\boldsymbol{n}\cdot\left[\boldsymbol{\psi}^\dagger(\boldsymbol{r})\boldsymbol{\hat\sigma}\boldsymbol{\psi}(\boldsymbol{r})\right],\label{eq:curr}
	\ee
	vanishes along the boundary, yielding 
	\be
	\psi_2[s(\varphi)]=i e^{i\alpha[s(\varphi)]}\psi_1([s(\varphi)].
	\label{BC}
	\ee
	Here, the local current is the expectation value of the current operator $\boldsymbol{\hat u}=\boldsymbol{\nabla}_{\boldsymbol{p}}\hat H_D=c\boldsymbol{\hat\sigma}$, $s(\varphi)$ is the arc-length parameter with $\varphi$ parameterizing the boundary,
	\be
	s(\varphi)=\int_0^\varphi\vert w^\prime(\tilde\varphi)\vert d\tilde\varphi,\, s\in\left[0,\mathcal{L}\right], ds=\vert w^\prime(\tilde\varphi)\vert d\tilde\varphi 
	\ee
with $w^\prime(\varphi)=\frac{dw(\varphi)}{d\varphi}$, $\mathcal{L}$ is the perimeter and $\alpha[s(\varphi)]$ is the angle of the outward-pointing normal vector $\boldsymbol{n}=\left(\cos\alpha[s(\varphi)],\sin\alpha[s(\varphi)]\right)$ at $w[s(\varphi)]$ with respect to the $x$ axis. 
Note that there are alternative BCs, that warrant self-adjointness of the Dirac Hamiltonian and zero outgoing current; cf. Refs.~\cite{Gaddah2018,Greiner1994}. In Ref.~\cite{Berry1987} only the ultra-relativistic limit, i.e. massless NBs were considered. Massive ones and the transition to the nonrelativistic limit were analyzed in Refs~\cite{Dietz2020,Zhang2021,Dietz2022}.

We computed the eigenvalues and eigenfunctions of the NRQB and NB based on a boundary-integral method (BIM). For the NRQB the corresponding BIE is given by~\cite{Baecker2003}
	\begin{align}
	&&\partial_n\psi[s(\varphi^\prime)]=\int_0^{2\pi}d\varphi\vert w^\prime(\varphi)\vert Q^{NRQB}(k;\varphi,\varphi^\prime)\partial_n\psi[s(\varphi)],\label{BIE_QB}
	\\
	&&Q^{NRQB}(k;\varphi,\varphi^\prime)=i\frac{k}{2}\cos\left[\alpha(\varphi^\prime)-\Xi(\varphi,\varphi^\prime)\right]H_1^{(1)}(k\rho),\nonumber
	\end{align}
	where $\Xi$ and $\rho$ are the phase and modulus of the distance vector $\boldsymbol{r}(\varphi,\varphi^\prime)$ between two points along the boundary, 
	\be
	e^{i\Xi(\varphi,\varphi^\prime)}=\frac{w(\varphi)-w(\varphi^\prime)}{\vert w(\varphi)-w(\varphi^\prime)\vert},\, \rho(\varphi,\varphi^\prime)=\vert w(\varphi)-w(\varphi^\prime)\vert .
	\ee
	Furthermore, $H_{m}^{(1)}(x)$ is the Hankel function of the first kind of order $m$.

	The BIE for the first spinor-eigenfunction of the NB is given by~\cite{Berry1987} 
	\bea
	\label{BIE_NB}
	&&\psi_1^\ast(\varphi^\prime)=\frac{ik}{4}\int_0^{2\pi}d\varphi\vert w^\prime(\varphi)\vert Q_1^{NB}(k;\varphi,\varphi^\prime)\psi_1^\ast(\varphi),\\
	&&Q_1^{NB}(k;\varphi,\varphi^\prime)=\nonumber
	\left\{e^{i\left[\alpha(\varphi^\prime)-\alpha(\varphi)\right]}-1\right\}H_0^{(1)}(k\rho)+\\
	&&\left\{e^{i\left[\Xi(\varphi,\varphi^\prime)-\alpha(\varphi)\right]}+e^{-i\left[\Xi(\varphi,\varphi^\prime)-\alpha(\varphi^\prime)\right]}\right\}\nonumber H_1^{(1)}(k\rho).
	\nonumber
	\eea
	The corresponding equations for $\psi_2^\ast(\varphi^\prime)$ and $Q_2^{NB}(k;\varphi,\varphi^\prime)$ are obtained by applying~\refeq{BC} to~\refeq{BIE_NB}.

	There are several essential differences between NRQBs and NBs, namely 
	\begin{itemize}

	\item{Time-reversal invariance}

	The Dirac Hamiltonian~\refeq{DE} is not time-reversal invariant. Therefore, if the spectral properties of a typical NRQB with the shape of a chaotic CB agree well with those of random matrices from the GOE, then those of the corresponding NB exhibit GUE statistics~\cite{Berry1987} if there is no mirror symmetry, otherwise GOE statistics. 

	\item{Mirror symmetries}

	The eigenfunctions of a NRQB whose shape has a mirror symmetry can be separated into symmetric and antisymmetric ones with respect to the symmetry axes, that is,  they fulfill either Neumann or Dirichlet BCs along these lines. This is not possible for NBs, because the BCs~\refeq{BC} do not comply with mirror reflection~\cite{Yupei2020}. Still, the Dirac Hamiltonian exhibits the properties $\hat H_{D}(-x,y)=\hat H^\ast_{D}(x,y)$ and $\hat H_{D}(x,-y)=\hat\sigma_z\hat H^\ast_{D}(x,y)\hat\sigma_z$. Accordingly, some eigenfunctions of a NB with mirror symmetry with respect to the $y$ or $x$ axes will have the property  $\left[\psi_1(-x,y),\psi_2(-x,y)\right]=\pm\left[\psi_1^\ast(x,y),\psi_2^\ast(x,y)\right]$ or $\left[\psi_1(x,-y),\psi_2(x,-y)\right]=\pm\left[\psi_1^\ast(x,y),-\psi_2^\ast(x,y)\right]$, respectively. 

	\item{Discrete rotational symmetry}

	The eigenstates of a NRQB, whose shape has a $Q$-fold rotational symmetry can be separated into $Q$ subspaces labeled by $l=0,\dots ,Q-1$ according to their transformation properties under the operator $\hat R^{\lambda}$ for a rotation by $\frac{2\pi}{Q}$~\cite{Robbins1989,Leyvraz1996,Keating1997,Joyner2012}
	\be
	\hat R^{\lambda}\psi^{(l)}_m(r,\varphi)=e^{il\frac{2\pi}{Q}\lambda}\psi^{(l)}_m(r,\varphi).
	\label{RotSym}
	\ee
	Similarly, the spinor components of the eigenstates of the corresponding NB can be classified according to their transformation properties under a rotation by $\frac{2\pi}{Q}$ into the $Q$ subspaces~\cite{Dietz2021,Zhang2021,Zhang2023}, however, they belong to different ones. If $\psi_{1,m}(\boldsymbol{r})$ belongs to the subspace $l$,
	\be
	\hat R\psi_{1,m}(\boldsymbol{r})=e^{il\frac{2\pi}{Q}}\psi_{1,m}(\boldsymbol{r}),\label{sympsi1}
	\ee
	then $\psi_{2,m}(\boldsymbol{r})$ belongs to the subspace $\tilde l =(l-1)$,
	\be
	\hat R\psi_{2,m}(\boldsymbol{r})=e^{i(l-1)\frac{2\pi}{Q}}\psi_{2,m}(\boldsymbol{r}),\label{sympsi2}
	\ee
	where $\tilde l=-1$ corresponds to $l=Q-1$. This implicates, that the eigenspinors themselves can not be classified according to their transformation properties under rotation by $\frac{2\pi}{Q}$. The intermixture of symmetry classes originates from the additional spin degree of freedom~\cite{Zhang2021,Dietz2021} and is a consequence of the BC~\refeq{BC}. It has, for example, consequences for the spectral properties of sectors of the circular NB. Namely, these do not have any common eigenstates with the circular NB and their spectral properties are not Poissonian, but intermediate~\cite{Yupei2022}. 

	\item{Chirality}

		For NRQBs, the length spectra, i.e., the Fourier transform of the fluctuating part of the spectral density $\rho^{fluc}(k)$ from wave number to length, exhibits peaks at the lengths of periodic orbits (POs) of the corresponding classical billiard (CB), whereas in the length spectra of the corresponding NB peaks at the lengths of POs with an odd number of reflections at the boundary are absent. These features have been shown in Ref.~\cite{Berry1987} to have their origin in the chirality of such POs. The connection between length spectra and POs was established within the periodic-orbit theory in the semiclassical limit $\hbar\to 0$, which was pioneered by Gutzwiller~\cite{Gutzwiller1971,Gutzwiller1990}. Within this semiclassical approach Gutzwiller derived for chaotic systems trace formulas for the fluctuating part of the spectral density of a quantum system in terms of a sum over the POs of the associated classical dynamic. Based on the semiclassical Einstein-Brillouin-Keller quantization~\cite{Einstein2017,Brillouin1926,Keller1958} Berry and Tabor derived a trace formula~\cite{Berry1977a} for integrable systems. In Refs.~\cite{Dietz2019} and~\cite{Dietz2020} corresponding trace formulas were derived based on the BIM for massive and massless NBs with shapes generating an integrable and chaotic dynamics, respectively, following the procedure developed in Refs.~\cite{Harayama1992,Sieber1997a,Sieber1998}, also for NBs whose shape exhibits a discrete rotational symmetry in Refs.~\cite{Zhang2021,Zhang2023c}. 

	Gutzwiller's trace formula for NRQBs reads   
	\be     
	\rho^{fluc}(k)=
	\frac{1}{\pi}\Re\sum_{\gamma_p}\mathcal{A}_{\gamma_p}e^{i\Theta_{\gamma_p}}\label{rhoNRQB},
	\ee
	and that for massless NBs is given by
	\begin{align}
	\rho^{fluc}(k)=&\frac{1}{\pi}\Re\sum_{\gamma_p}(-1)^p\cos\left(\Phi_{\gamma_p}\right)\cos\left(p\frac{\pi}{2}\right)\label{rhoNB}\\
		&\times\mathcal{A}_{\gamma_p}e^{i\Theta_{\gamma_p}}\nonumber.
	\end{align}
	The sum is over periodic orbits $\gamma_p$ with $p$ reflections at the boundary which are $p$ periodic, and $\mathcal{A}_{\gamma_p}$ and $\Theta_{\gamma_p}$ denote amplitudes and phases, 
	\begin{align}
	&\mathcal{A}_{\gamma_p}=\frac{l^{p}_{\rm PO}}{r^{p}_{\rm PO}\sqrt{\vert{\rm Tr}M_{\rm PO}^{p}-2\vert}},\nonumber\\
	&\Theta_{\gamma_p}=kl_{\rm PO}^{p}-\frac{\pi}{2}\mu_{\rm PO}^{p}.\label{Phase_Ampl}
	\end{align}
	Here $l^{p}_{\rm PO},\, \mu_{\rm PO}^{p},\ M_{\rm PO}^{p}$ denote the length, Maslov index, and monodromy matrix of the periodic orbit, and $r^{p}_{\rm PO}$ the number of repetitions of the primitive periodic orbit. For the NB only periodic orbits with an even number of reflections $p$~\cite{Bolte1999,Wurm2011} contribute to the trace formula. This feature originates from the chirality property and the additional spin degree of freedom, that is, the vectorial character of the Dirac equation~\cite{Balian1977,Dembowski2002,Wurm2011}.
	\end{itemize}

	\subsection{Properties of the eigenstates of the stadium NRQB and NB~\label{SpectrNRQBNB}}
	We employed the BIEs~(\ref{BIE_QB}) and~(\ref{BIE_NB}) to compute 7500 eigenvalues for the full- and quarter-stadium NRQB and NB, respectively. This is sufficient, because the objective is to compare the properties of the eigenstates of the NRQB and NB to those of the corresponding GB and HGB, where only several hundreds of eigenstates are available. For the full stadium we exploited its twofold symmetry and computed the eigenstates belonging to the symmetry classes $l=0$ and $l=1$ separately~\cite{Zhang2021}, thereby avoiding degeneracies in the eigenvalue spectrum of the NRQB and near-degeneracies in that of the NB. Furthermore, this reduces computation time considerably. 

\begin{figure}
	\centering
        \begin{tabular}{@{}c@{}}
        \begin{tabular}{@{}c@{}}
        \includegraphics[width=.42\linewidth]{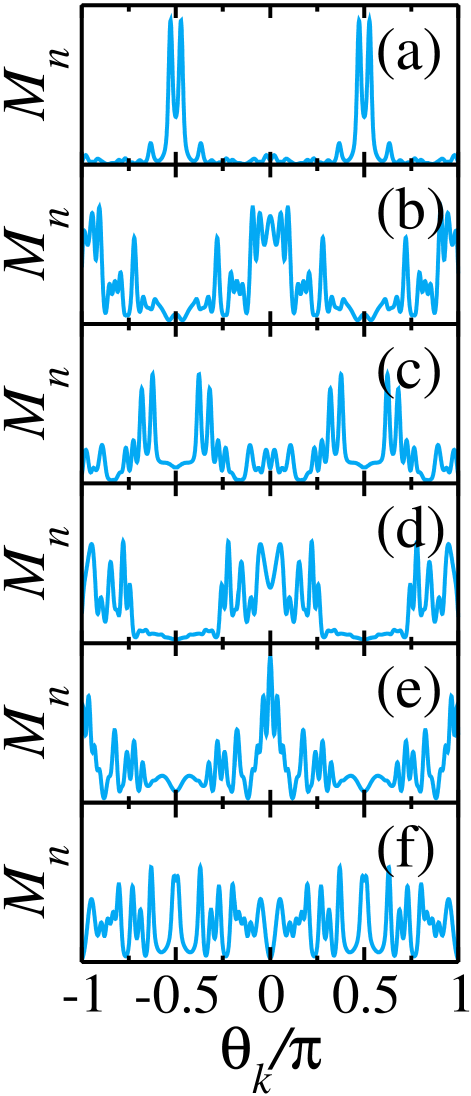}
        \end{tabular}
        \begin{tabular}{@{}c@{}}
        \includegraphics[width=.355\linewidth]{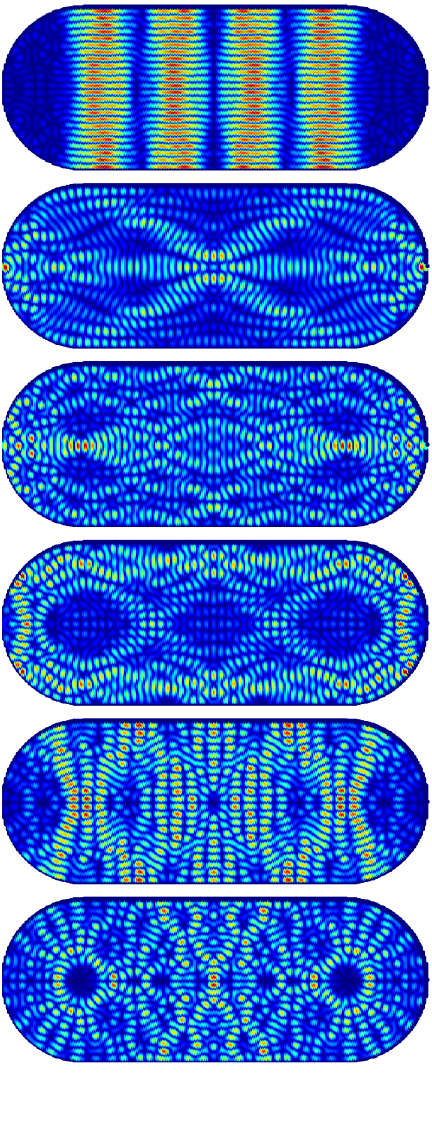} \\[\abovecaptionskip]
        \end{tabular}
	\begin{tabular}{@{}c@{}}
        \includegraphics[width=.075\linewidth]{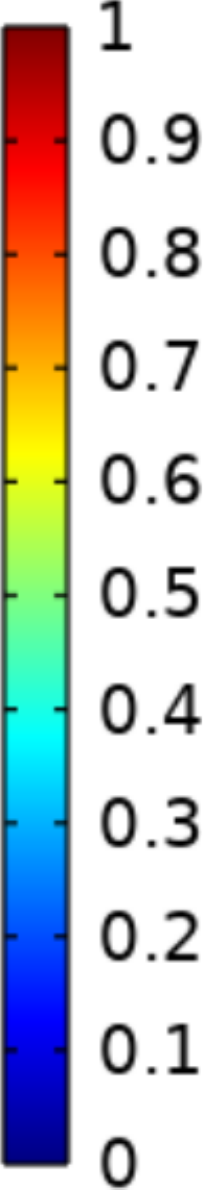} \\[\abovecaptionskip]
        \end{tabular}
        \end{tabular}
		\caption{On-shell momentum distribution of the first eigenspinor component $\Psi_{1,m}(\boldsymbol{r})$ of the NRQB versus momentum direction $\theta_k$ for (a) $m=1048$, (b) $m=1102$,  (c) $m=1020$, (d) $m=1005$, (e) $m=1005$, (f) $m=1040$. To the right is exhibited the modulus of the corresponding wave-functions $\psi_m(\boldsymbol{r})\vert$.}\label{Mom_QB} 
	\end{figure}

\begin{figure}
        \begin{tabular}{@{}c@{}}
        \begin{tabular}{@{}c@{}}
        \includegraphics[width=.42\linewidth]{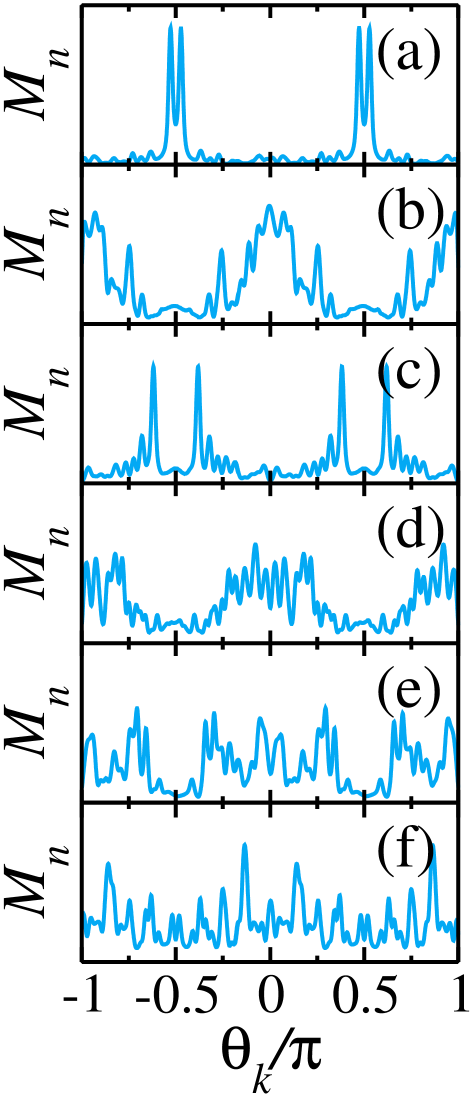}
        \end{tabular}
        \begin{tabular}{@{}c@{}}
        \includegraphics[width=.355\linewidth]{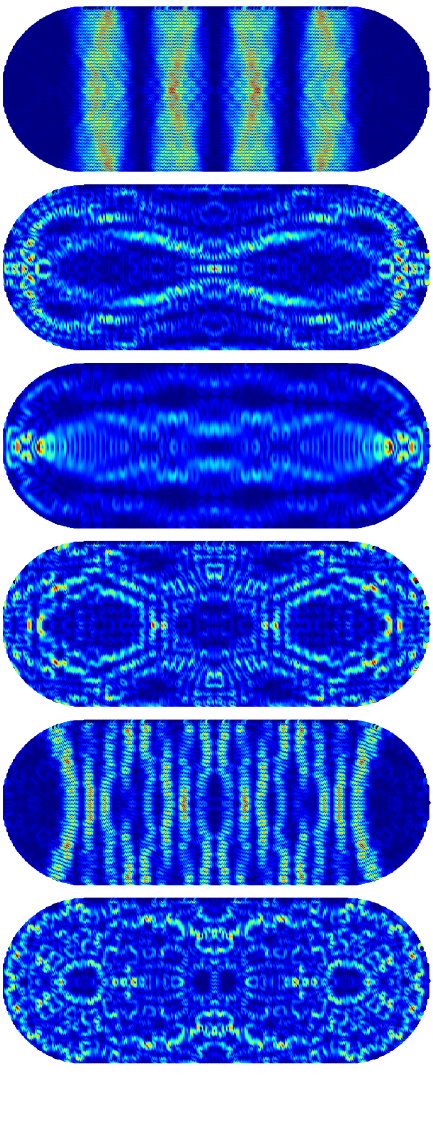} \\[\abovecaptionskip]
        \end{tabular}
        \end{tabular}
		\caption{On-shell momentum distribution of the first eigenspinor component $\Psi_{1,m}(\boldsymbol{r})$ of the NB versus momentum direction $\theta_k$ for (a) $m=1136$, (b) $m=482$, (c) $m=1001$, (d) $m=1141$, (e) $m=1119$, (f) $m=1130$. To the right are exhibited the distributions of the corresponding local current $\vert\boldsymbol{u}_m(\boldsymbol{r})\vert$.}\label{Mom_NB}
        \end{figure}

\begin{figure}
	\centering
	\includegraphics[width=.45\linewidth]{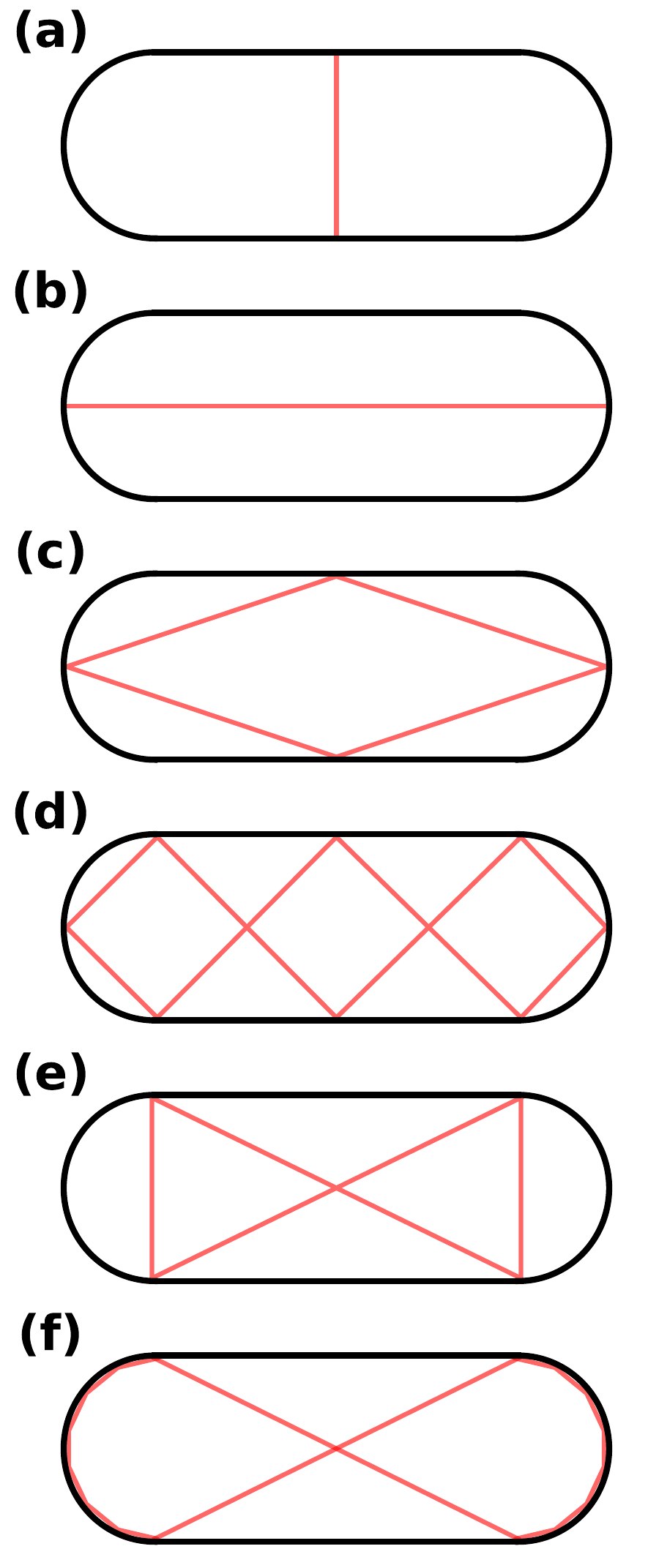} \\[\abovecaptionskip]
		\caption{Enhanced localization is observed along orbits from the family of the (a) BBO [cf. Figs.~\ref{Mom_QB} and~\ref{Mom_NB} (a)], (b) the diameter orbit [cf. Figs.~\ref{Mom_QB} and~\ref{Mom_NB} (c)], (c) the diamond orbit reflected at the centers of the semicircular parts and at the centers of the straight parts of the boundary and (e) orbits that, like the orbit (c) look similar to those of a rectangular billiard with side lengths $(2L)\times (2L+2r_0)$, and are reflected at the centers of the semicircular parts and at the straight parts of the boundary [cf. Figs.~\ref{Mom_QB} and~\ref{Mom_NB} (d)] (e) the bow-tie orbit constructed from reflections at opposite and diagonal corners, respectively [cf. Figs.~\ref{Mom_QB} and~\ref{Mom_NB} (e)] and (f) edge orbits composed of whispering gallery orbits of the semicircular parts and diagonal orbits that connect two corners [cf. Figs.~\ref{Mom_QB} and~\ref{Mom_NB} (b)].} 
	\label{Scars}
	\end{figure}

	In Figs.~\ref{Mom_QB} and~\ref{Mom_NB} are shown typical intensity distributions of the wave functions and local-current, respectively. The distributions shown in (a)-(e) exhibit enhancement along periodic orbits. Generally, we observe such a scarring along periodic orbits, that are members of the families presented in~\reffig{Scars}. Shown are examples from the family of (a) the BBOs bouncing back and forth between the straight parts of the boundary, (b) isolated diameter orbits bouncing back and forth between the centers of the semicircular parts, (c) the diamond orbit and (d) another orbit reflected at the centers of the circular parts and at the straight parts like in a rectangular billiard with side lengths $(2L)\times (2L+2r_0)$, (f) bow-tie orbits reflected at opposite and diagonal corners, and (f) edge orbits composed of whispering-gallery orbits along the semicircular parts and diagonal orbits connecting two corners or proceeding along the straight parts of the boundary. 

	\begin{figure}
	\centering
	\includegraphics[width=\linewidth]{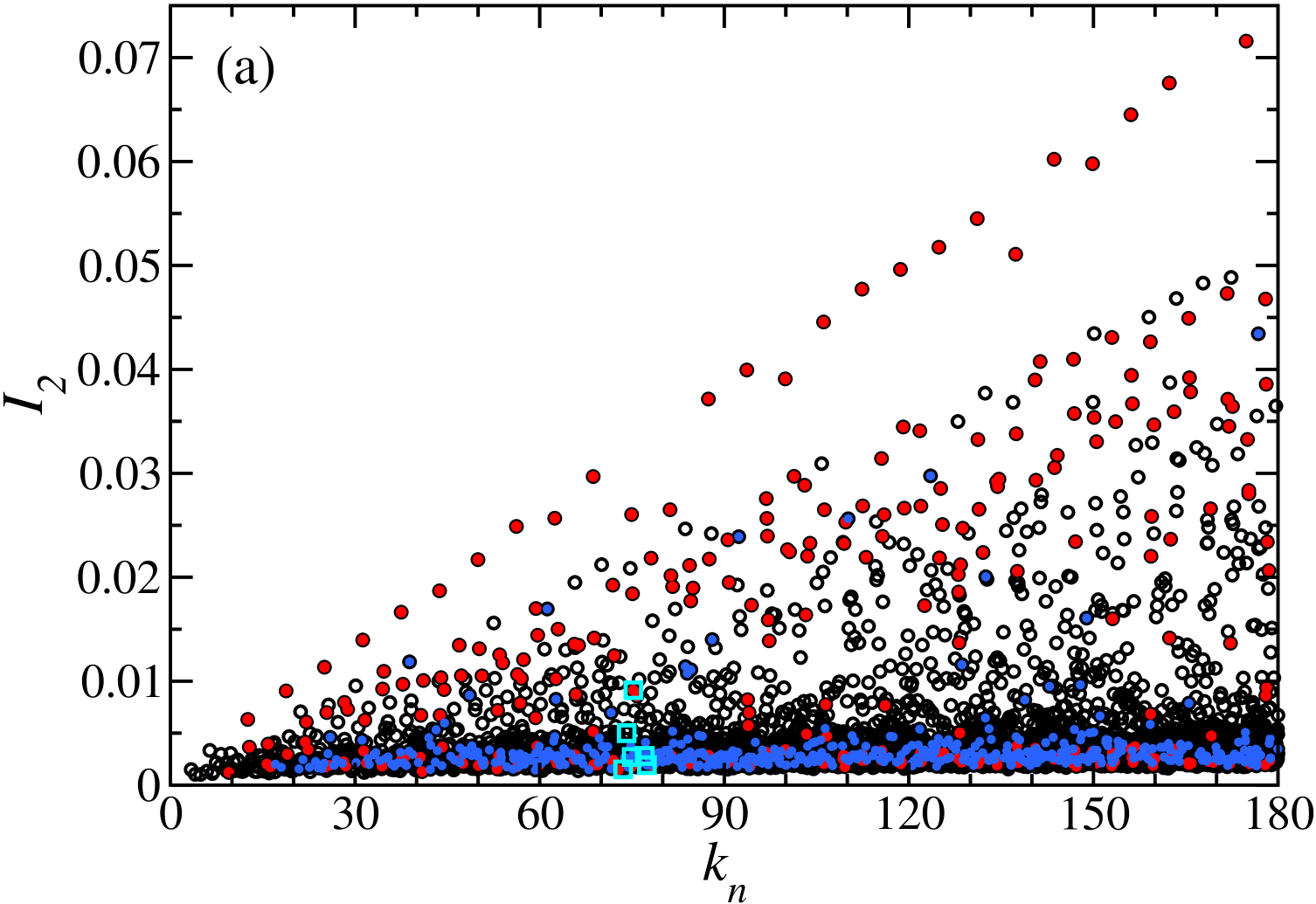}
	\includegraphics[width=\linewidth]{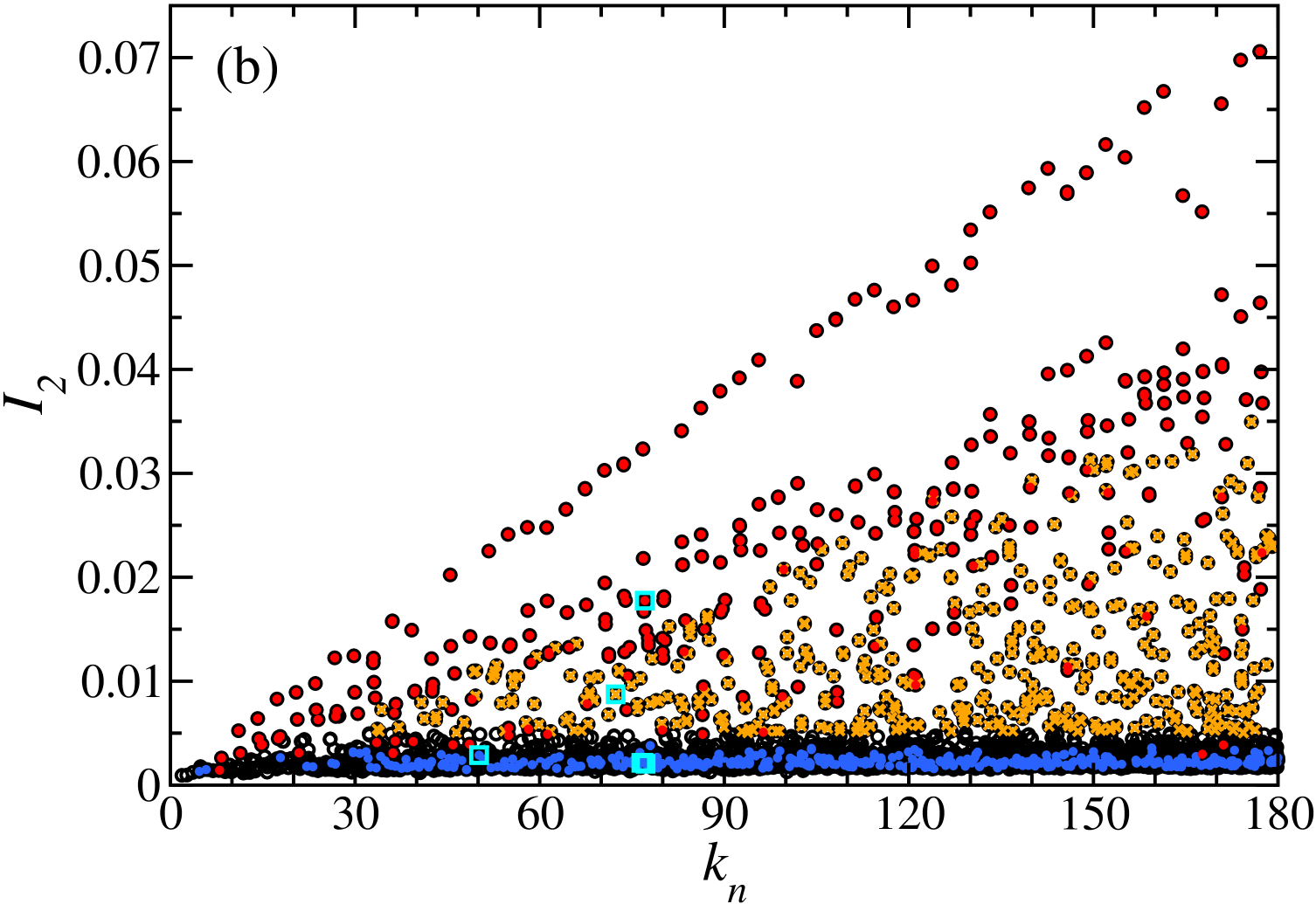}
		\caption{(a) Participation number of the on-shell momentum distribution of eigenstates $\Psi_m(\boldsymbol{r})$ of the NRQB. Red dots show the participation numbers for the BBOs and blue dots those of edge contributions from states exhibiting enhanced localization along whispering gallery modes of the semicircular parts of the boundary and are either also localized along the straight parts or reflected at the corners joining these parts. Examples for such wave functions are shown in Fig.~\ref{Mom_QB} (b) - (d), respectively. The participation numbers of the momentum distributions shown in ~\reffig{Mom_QB} are marked by cyan squares.
		(b) Same as (a) for the first eigenspinor component $\Psi_{1,m}(\boldsymbol{r})$ of the NB, with examples shown in Fig.~\ref{Mom_NB} (b) - (d), respectively. Orange crosses result from wave functions that are localized along almost BBOs that are reflected at an angle close to $90^\circ$ along the straight part and leak into the semicircular parts or are localized along orbits from the family of the bow-tie orbit~\reffig{Scars} (e), which are predominantly localized in the rectangular part. Examples are shown in Figs.~\ref{Mom_QB} and~\ref{Mom_NB} (e). The participation numbers of the momentum distributions shown in ~\reffig{Mom_NB} are marked by cyan squares.}
	\label{IPR_Mom_QB_NB}
	\end{figure}
	\begin{figure}
        \centering
        \includegraphics[width=\linewidth]{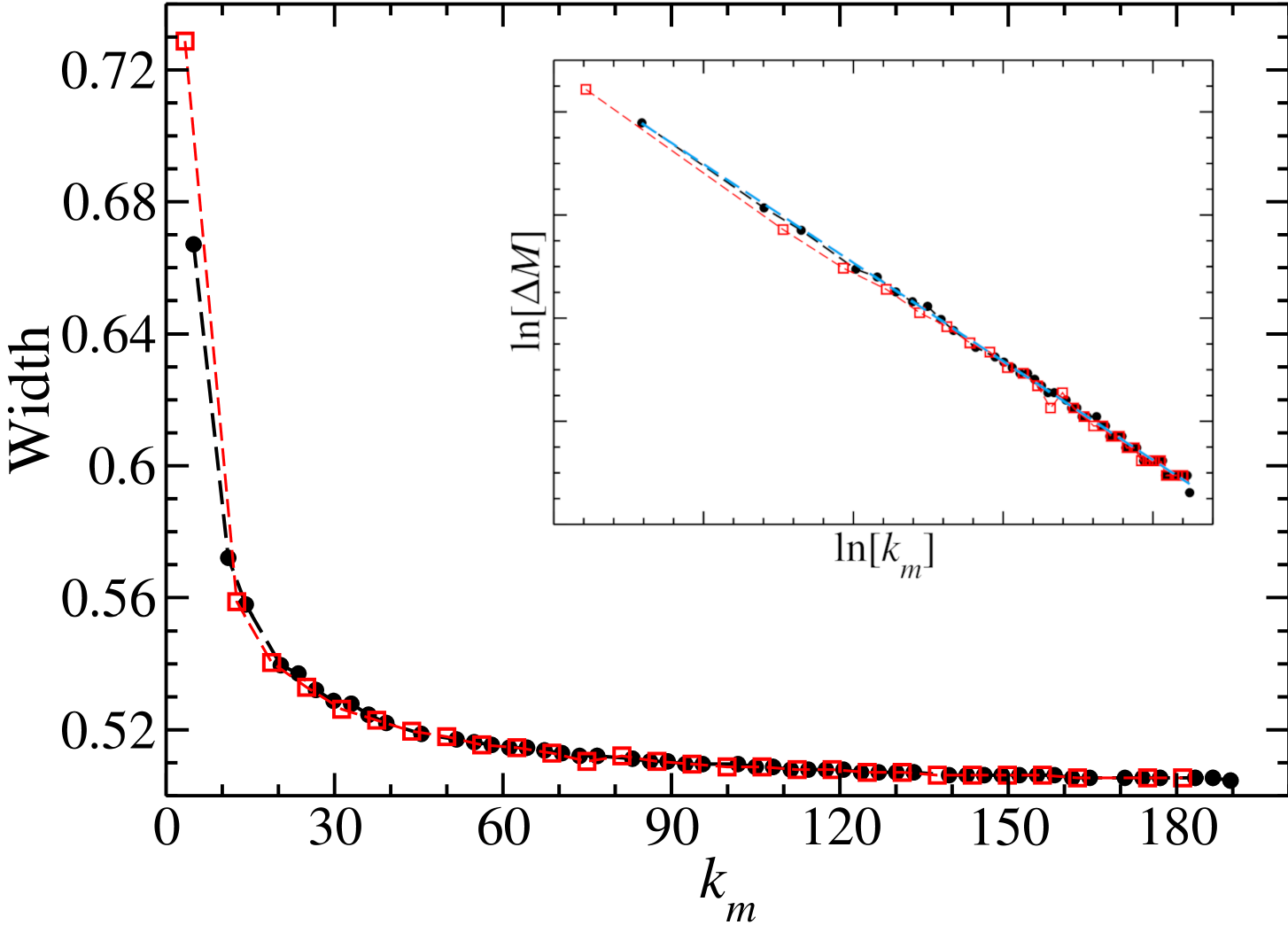}
		\caption{Widths of the momentum distributions of the two-bounces BBO [cf.~\reffig{Scars} (a)], corresponding to the largest values of the particiption numbers which form a straight line (cf.~\reffig{IPR_Mom_QB_NB}) for the NRQB (red squares) and the NB (black dots). They decrease algebraically with $k$ as illustrated in the log-log plot in the inset. A fit of a straight line to the data points reveals that the widths decay $\propto k^{-1}$ for the NRQB and NB.}
        \label{Width_Mom_QB_NB}
        \end{figure}
To identify such scarred wave functions and spinor functions we computed, inspired by Refs.~\cite{Bogomolny2006,Bogomolny2021a}, momentum distributions~\cite{Baecker1999}, i.e. their Fourier transforms from coordinate space $(x,y)$ to momentum space $(k_x,k_y)$ to identify such scarred wave functions and spinor functions. Momentum distributions are localized on the energy shell, that is, at values $k=k_m=\sqrt{k_{m,x}^2+k_{m,y}^2}$. Therefore, we restricted to it, 
	\be
	\label{Mom_Distr}
	\mathcal{M}_m(\theta_k)=\int_\Omega\Psi_m(x,y)e^{-ixk_{m,x}-iy\sqrt{k^2_m-k_{m,x}^2}}dxdy.
	\nonumber\ee
	It has been well established by now that the wave functions of typical quantum systems with a chaotic classical dynamics are well described by a superposition of random plane waves with fixed wavenumber $k=k_m$ and random directions $\theta_k$, whereas those of a typical quantum system with integrable classical dynamics are described by a superposition of plane waves with finite number of well-defined directions. The momentum distribution provides information on the preferred directions $\theta_k=\arctan(k_{m,y}/k_{m,x})$ of the plane waves that form the eigenmode~\cite{Bogomolny2021a}. 

In Figs.~\ref{Mom_QB} and~\ref{Mom_NB} are shown examples of the on-shell momentum-distributions~\refeq{Mom_Distr} for typical wave functions together with them. Those of the NRQB are symmetric with respect to $\theta_k=0$, whereas those of the NB do not have this symmetry. This is attributed to the BCs~\refeq{BC} which lead to a unidirectionality of the local current along the boundary and chirality. However, they exhibit a $\pi$ periodicity, originating from the two-fold symmetry of the stadium. In both figures, in (a) it is peaked around $\theta_k/\pi=\pm 1/2$ as expected for BBOs, in (b) and (d) it exhibits peaks around $\theta_k/\pi=0\, (\pm 1)$ as for whispering-gallery like orbits. Furthermore, in (b) it has sharp peaks at $\theta_k/\pi\simeq \pm 0.176\, (\pm 0.824),\pm 0.255\, (\pm 0.745)$, corresponding to the angles of the straight lines connecting diagonal corners and at $\theta_k/\pi\simeq\pm 1/4\, (\pm 3/4)$, respectively. In c) it is localized at $theta_k/\pi\simeq \pm 0.38\, (\pm 0.62)$ corresponding to trajectories reflected at the semicircular parts close to their centers and e) it is localized close to $\pm 1/3\, (\pm 2/3$ and in (f) it extends over the whole $\theta_k$ range, as in typical wave functions of chaotic systems. We used these features of the momentum distribution to identify eigenstates of the NRQB and NB that are scarred along BBOs and almost BBOs that leak into the semicircular parts like in Figs.~\ref{Mom_QB} and~\ref{Mom_NB} (a) and (e), and edge orbits. To quantify the localization properties of the momentum distributions reflected in the peak structure, we generalized the participation-number, which serves in this context as a measure for wave functions, and define them accordingly as 
	\be
	\mathcal{I}_2=\frac{\sum_{i=1}^{2N_p+1}\left\vert\mathcal{M}_m\left(\theta_{k,i}\right)\right\vert^4}{\left(\sum_{i=1}^{2N_p+1}\vert\mathcal{M}_m\left(\theta_{k,i}\right)\vert^2\right)^2}.\label{IPR}
	\ee
The momentum distributions and participation numbers are computed for $2N_p+1$ discrete values of $\theta_{k,i}\in [-\pi,\pi],\,i=1,2N_p+1$. Note that in distinction to wave functions the momentum distribution is not necessarily normalized to unity, so that the denominator needs to be included. The resulting participation numbers are shown in~\reffig{IPR_Mom_QB_NB}. In both the NRQB and NB, the largest values correspond to BBOs. Furthermore, for the NB also the almost BBOs that leak into the semicircular parts or are localized on orbits from the family of the bow-tie orbit, shown as yellow crosses, have larger participation numbers than all other eigenstates, whereas for the NRQB also other orbits have values of similar size. For the edge orbits, marked by blue dots, the values are comparably small (blue dots) in NRQB and NB. We conclude from these results that the enhancement of localization along BBOs, almost BBOs and orbits from the family of the bow-tie orbits is strongest for the NBs, whereas in the NRQB wave functions that are localized along different orbits may exhibit stronger localization than such orbits. This feature is attributed to the differing BCs. We should mention, that the first and second spinor component of the eigenspinors of the NB, that are localized along BBOs, have the property $\psi_1(x,y)=\psi_1(x,-y)$, $\psi_1(x,y)=\psi_1^\ast(x,-y)$ and $\psi_2(x,y)=\psi_2(x,-y)$, $\psi_2(x,y)=-\psi_2^\ast(x,-y)$, respectively, resulting from the mirror symmetries with respect to the $x$ and $y$ axes, implying that $\psi_1(\boldsymbol{r})$ and $\psi_2(\boldsymbol{r})$ are real for such scarred spinors.          

In both the NB and NRQB the largest participation numbers correspond to BBOs with two bounces as depicted in~\reffig{Scars} (a). They lie on a straight line. To further investigate the origin of this behavior we analyzed the widths of the associated distributions, which are shown in~\reffig{Width_Mom_QB_NB}. The resulting curves for the NB and NRQB lie on top of each other. The corresponding log-log plot, shown in the inset, reveals that the widths decay with $k^{-1}$.    

	\begin{figure}
	\centering
	\begin{tabular}{@{}c@{}}
        \begin{tabular}{@{}c@{}}
	\includegraphics[width=.43\linewidth]{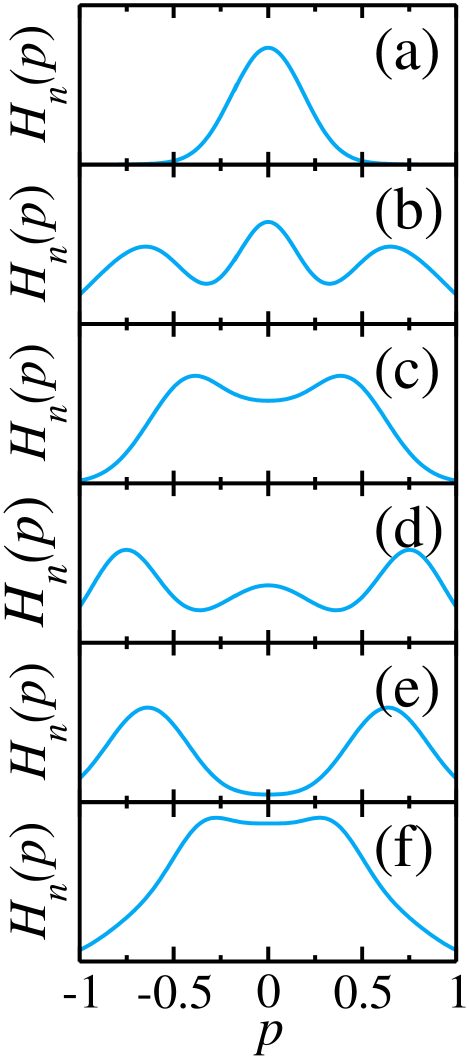}
        \end{tabular}
        \begin{tabular}{@{}c@{}}
	\includegraphics[width=.38\linewidth]{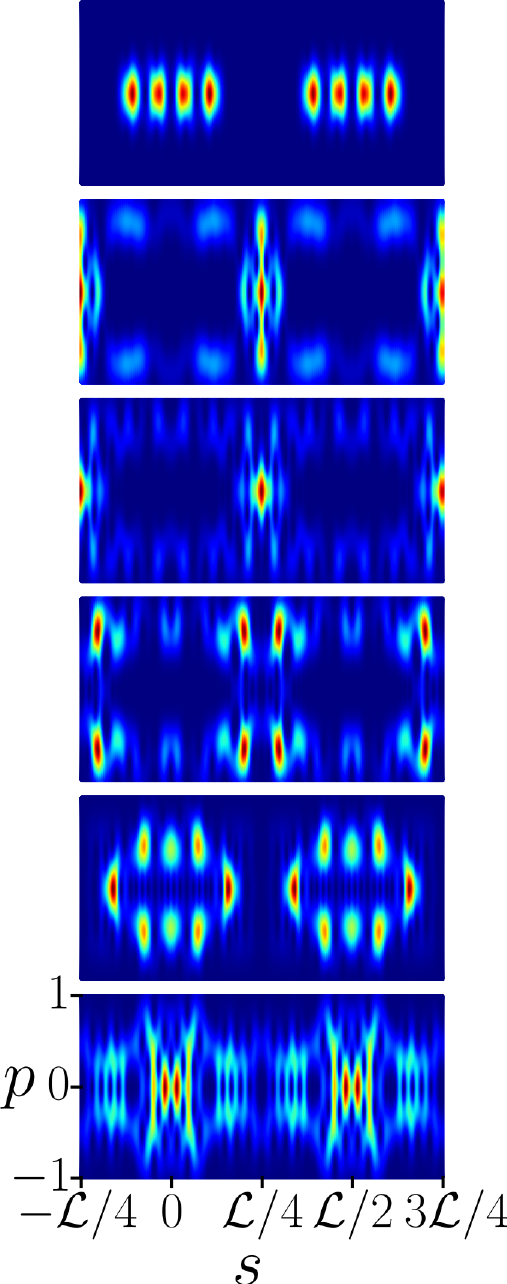}
        \end{tabular}
        \end{tabular}
		\caption{Projection of the Husimi functions onto the momentum ($p$) axis for eigenstates $\Psi_m(\varphi)$ of the NRQB for (a) $m=1048$, (b) $m=1102$, (c) $m=1020$, (d) $m=1005$, (e) $m=1112$, (f) $m=1040$. To the right are shown the corresponding Husimi functions $H_m(s,p)$ in the $(s,p)$ plane. The arclength $s$ is measured with respect to the center point of the lower straight part of the boundary and increases in counterclockwise direction along the boundary. The values $s=\mathcal{L}/4,\mathcal{L}/2,3\mathcal{L}/4$ ($-\mathcal{L}/4$) correspond to the centers of the right semicircular part, the upper straight part and the left semicircular part, respectively.}\label{Hus_QB}
	\end{figure}
	\begin{figure}
\begin{tabular}{@{}c@{}}
        \begin{tabular}{@{}c@{}}
        \includegraphics[width=.43\linewidth]{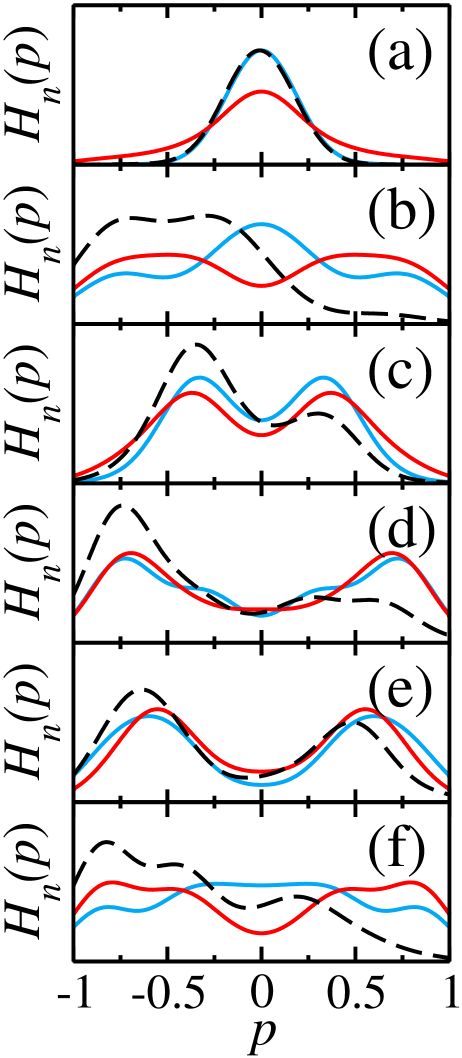}
        \end{tabular}
        \begin{tabular}{@{}c@{}}
        \includegraphics[width=.38\linewidth]{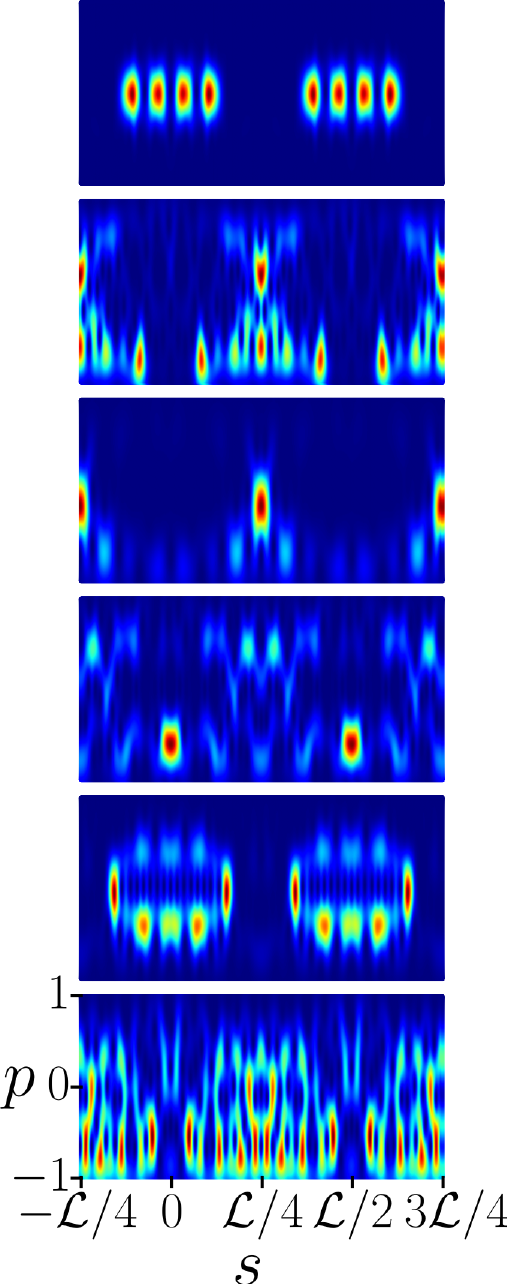} 
        \end{tabular}
        \end{tabular}
		\caption{Projection of the Husimi functions onto the momentum ($p$) axis for the first eigenspinor component $\Psi_{1,m}(\varphi)$ of the NB for (a) $m=1136$,  (b) $m=482$, (c) $m=1001$, (d) $m=1141$, (e) $m=1119$, (f) $m=1130$. Shown are the absolute values of the real part (blue solid lines), imaginary part (red solid lines) and the complete, complex-valued projection of the Husimi function (black dashed lines). The former two are symmetric with respect to $p=0$, the latter one is asymmetric. To the right are shown the corresponding Husimi functions $H_m(s,p)$ in the $(s,p)$ plane. The arclength $s$ is measured with respect to the center point of the lower straight part of the boundary and increases in counterclockwise direction along the boundary. The values $s=\mathcal{L}/4,\mathcal{L}/2,3\mathcal{L}/4$ ($-\mathcal{L}/4$) correspond to the centers of the right semicircular part, the upper straight part and the left semicircular part, respectively.}
\label{Hus_NB}
\end{figure}
The momentum distribution provides information concerning preferred directions of the waves propagating through the billiard, whereas Husimi functions give momentum directions at the boundary of trajectories along which the corresponding wave function is localized~\cite{Baecker2004}. They are defined in classical phase space~\cite{Husimi1940,LesHouches1989} and are often referred to as quantum Poincar\'e-surface-of-section for NRQBs. One of the authors demonstrated in Refs.~\cite{Yupei2022,Dietz2022,Dietz2023}, that for NBs Husimi functions provide similar information and thus may serve as a suitable tool to obtain information on the semiclassical limit. Following Ref.~\cite{Baecker2004} we define Husimi functions in terms of Poincar\'e-Birkhoff coordinates, given by the arc-length parameter $s$ at the point of impact and $p=\sin\chi(s)$ with $\chi(s)$ denoting the angle between the particle trajectory and the outward normal to the boundary at $s$, as the projection of the normal derivative of the boundary wave function $\psi(s)$, $\partial_n\psi(s)\vert_{s\in\partial\Omega}$ onto a coherent state~\cite{Baecker2004} which is localized at $\partial\Omega$ and periodic with period $\mathcal{L}$,
	\begin{align}
	&H_m(s,p)=\label{Hus}\\
	&\frac{1}{2\pi k_m}\frac{1}{\int_0^\mathcal{L} ds^\prime\left\vert\partial_{n^\prime}\psi(s^\prime)\right\vert^2}
	\left\vert\int_0^\mathcal{L} ds^\prime\partial_{n^\prime}\psi(s^\prime) C^\delta_{(s^\prime,p)}(s^\prime;k_m)\right\vert^2.
	\nonumber\\
	&C^\delta_{(s,p)}(s^\prime;k_m)=\nonumber
	\left(\frac{k_m}{\pi\delta^2}\right)^{1/4}\\
	&
	\times\sum_{m=-\infty}^\infty\exp\left(ipk_m\left(s^\prime-s+m\mathcal{L}\right)-\frac{k_m}{2\delta^2}\left(s^\prime-s+m\mathcal{L}\right)^2\right).
	\nonumber
	\end{align}
	Here $n^\prime=n(s^\prime)$ is the outward normal at $s^\prime$ and $\delta$ controls the resolution. For NBs $\psi(s)$ is replaced either by the first or the second eigenspinor component. Figures~\ref{Hus_QB} and~\ref{Hus_NB} exhibit for the NRQB and NB Husimi functions corresponding to the wave functions and local currents shown in Figs.~\ref{Mom_QB} and~\ref{Mom_NB}, respectively. The Husimi functions of the NB are asymmetric with respect to the $p=0$ axis. This is attributed to the BCs~\refeq{BC}, which induce a unidrectional current and chirality along the boundary $\partial\Omega$, and thus the observed discrepancies between the clockwise and counterclockwise modes, which is absent in the NRQB. 

The starting point of the arclength parameter $s$ is chosen at the center point of the lower straight part of the boundary of the stadium and increases counterclockwise. The values $s=-\mathcal{L}/4,0,\mathcal{L}/4,\mathcal{L}/2,3\mathcal{L}/4$ correspond the centers of the left semicircular part, the lower straight part, the right semicircular part, the upper straight part and the left semicircular part, respectively. For the BBOs the projection of $H_m(s,p)$ onto the $p$ axis, $H_m(p)=\frac{1}{\mathcal{L}}\int_0^{\mathcal{L}}dsH_m(s,p)$ is peaked at $p=0$ at the $s$ values of the stripes observed in Figs.~\ref{Hus_QB} and~\ref{Hus_NB} (a). For the diameter orbit it is peaked at $p=0$ at the points of impact at $s=-\mathcal{L}/4,\mathcal{L}/4,3\mathcal{L}/4$ shown in Figs.~\ref{Hus_QB} and~\ref{Hus_NB} (c). For intensity distributions localized along edge orbits like in Figs.~\ref{Hus_QB} and~\ref{Hus_NB} (b) and (d), it is nonzero in the regions of the semicircular parts at $\vert p\vert\simeq \pm 1$, as is the case for whispering-gallery modes. In Figs.~\ref{Hus_QB} and~\ref{Hus_NB} (e) it is peaked at values of $s$ where the corresponding wave function or local current is maximal in Figs.~\ref{Mom_QB} and~\ref{Mom_NB} (e) and values of $p$ corresponding to the directions of the orbits on which these distributions are localized. 

Thus, the features of the Husimi functions are clearly distinct for the different types of periodic orbits shown in~\reffig{Scars}. We, actually used a combination of the Husimi functions, momentum distributions and their participation numbers to identify wave functions and spinor funcions localized along BBOs, almost BBOs, bow-tie like orbits and edge orbits. In the latter case we did not distinguish between egde states of the type ~\reffig{Scars} (f) and whispering-gallery mode like edge orbits which evolve along the whole boundary. These are marked by red and blue dots and orange crosses in~\reffig{IPR_Mom_QB_NB}, respectively.    

\begin{figure}
	\centering
	\includegraphics[width=.9\linewidth]{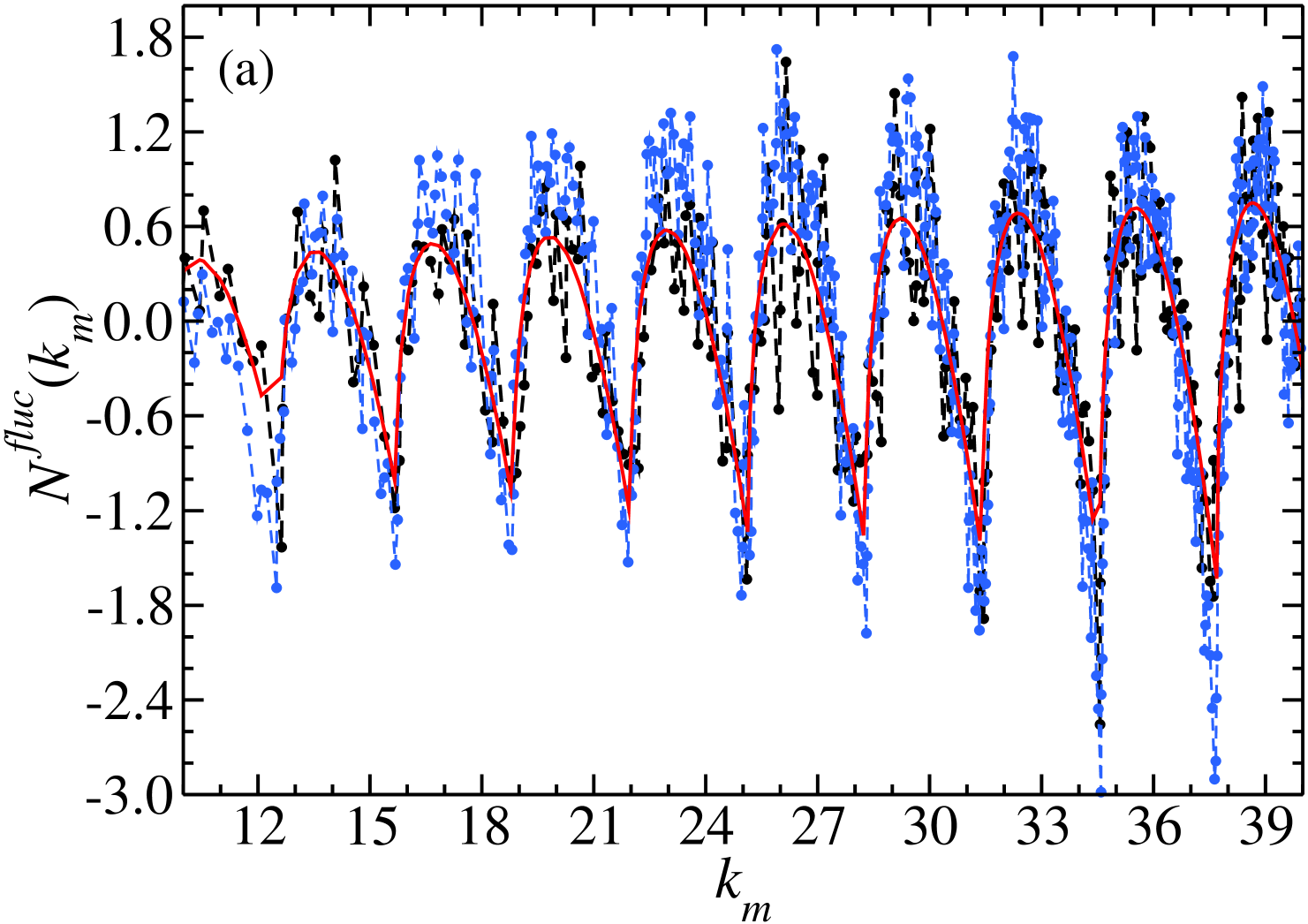}
	\includegraphics[width=.9\linewidth]{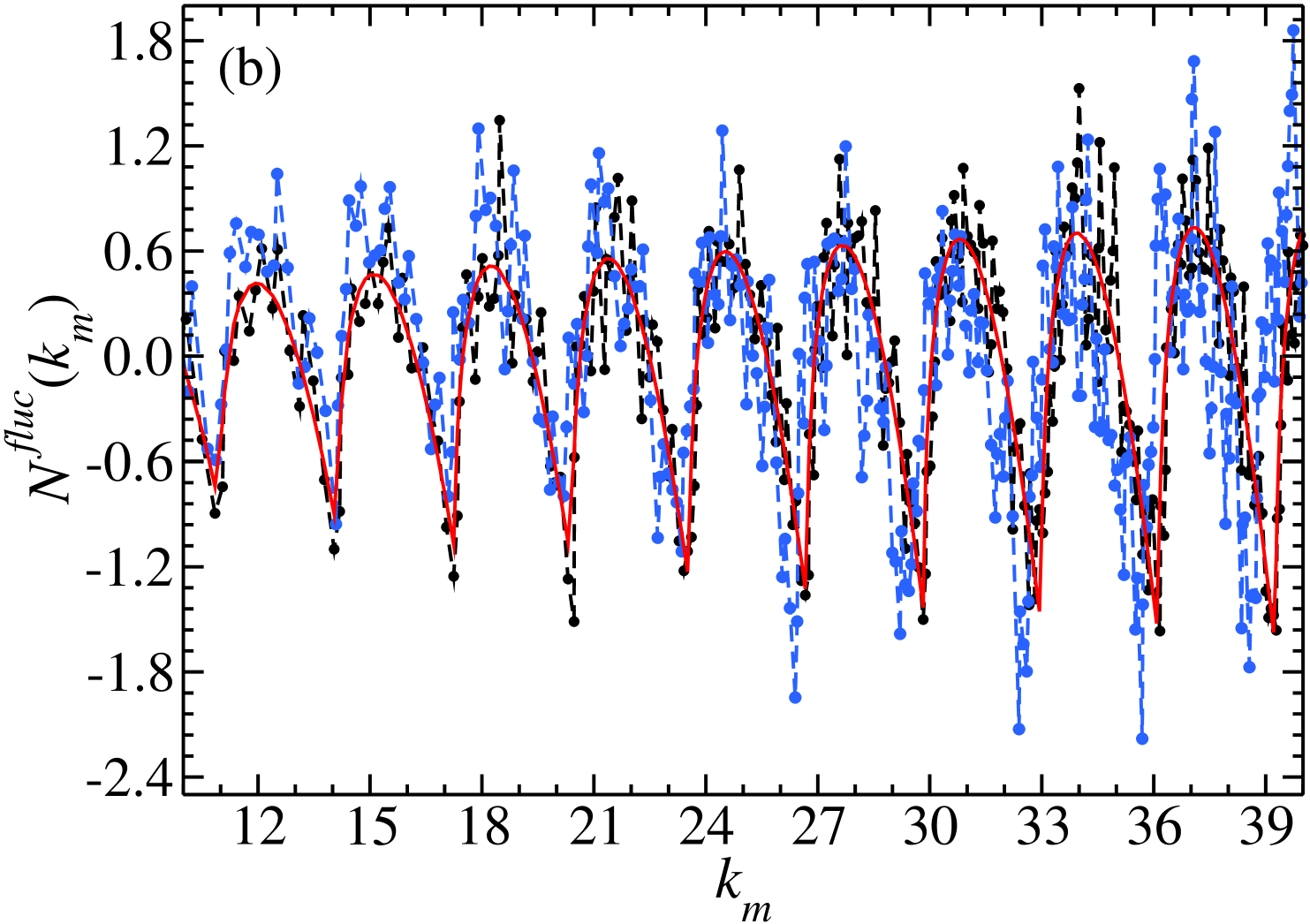}
	\caption{(a) Fluctuating part of the integrated spectral density of the quarter-stadium NRQB (black dots) compared to that of the BBOs deduced from~\refeq{NbboNRQB} (red curve) and of the GB (blue dots). 
	(b) Fluctuating part of the integrated spectral density of the quarter-stadium NB (black dots) compared to that of the BBOs deduced from~\refeq{NbboNB} (red curve) and of the eigenvalues of the HGB (blue dots).}
	\label{Nfluc}
	\end{figure}

\begin{figure}
	\centering
	\includegraphics[width=\linewidth]{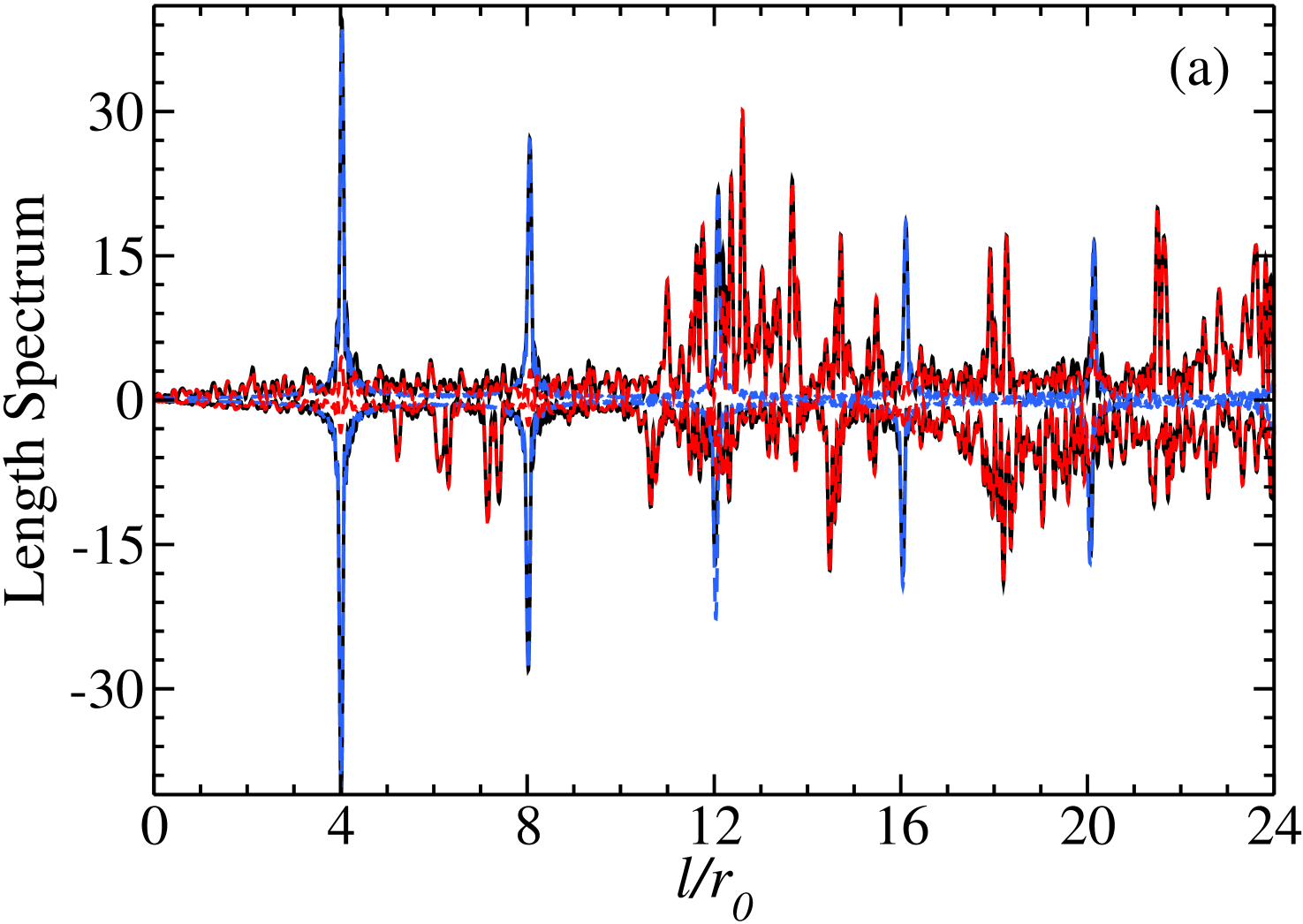}
	\includegraphics[width=\linewidth]{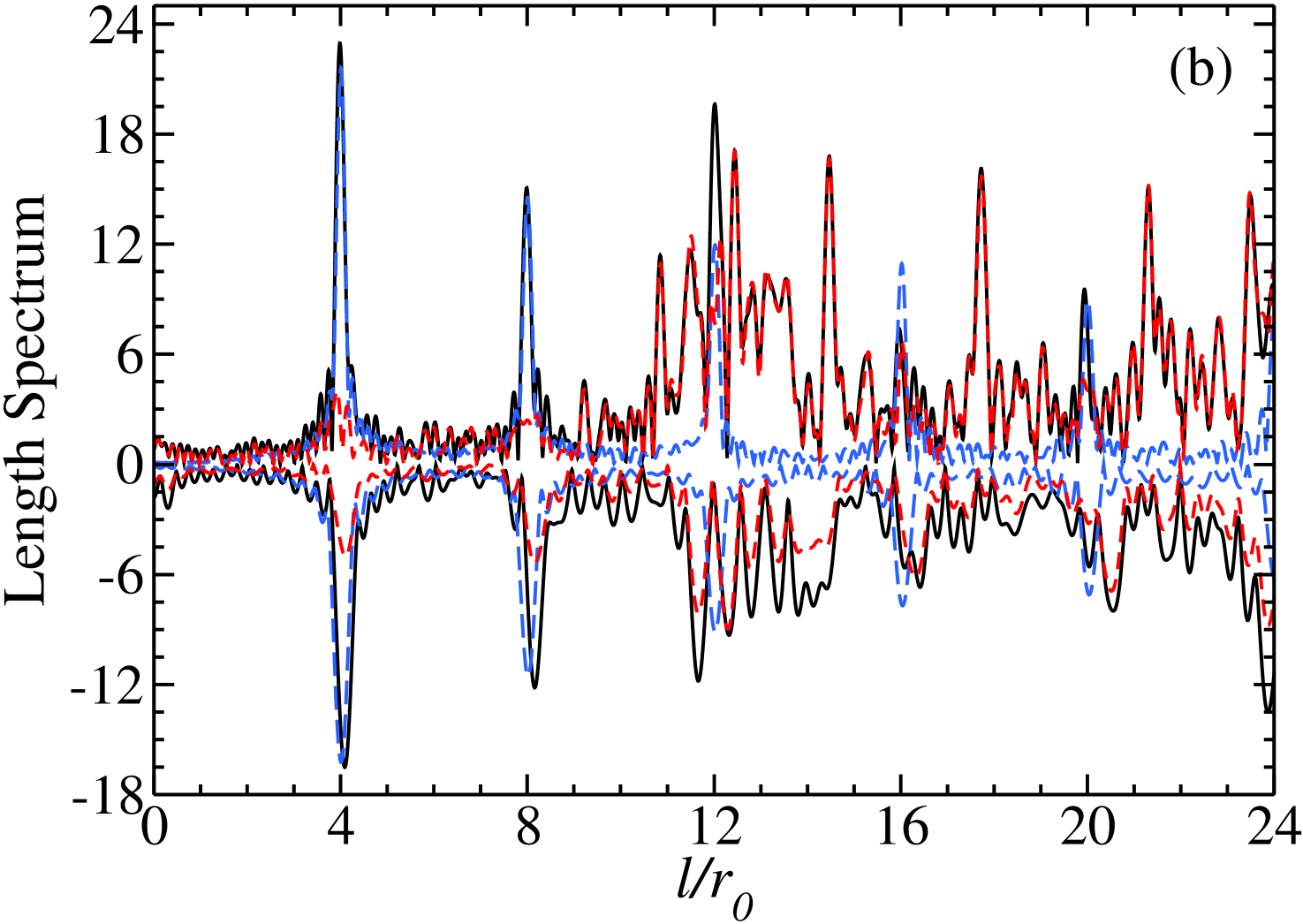}
		\caption{(a) Length spectrum (black) of the full stadium (NRQB) compared to that of the NB. The latter was multiplied with $(-1)$. Turquoise dashed lines show those of the BBOs and red dashed lines are obtained after extracting their contributions from $\mathcal{N}^{fluc}(k_m)$. (b) Same as (a) for the HGB around the band edge compared to that around the Dirac point.
	}\label{FFT_Full}
	\end{figure}

\begin{figure}
	\centering
	\includegraphics[width=\linewidth]{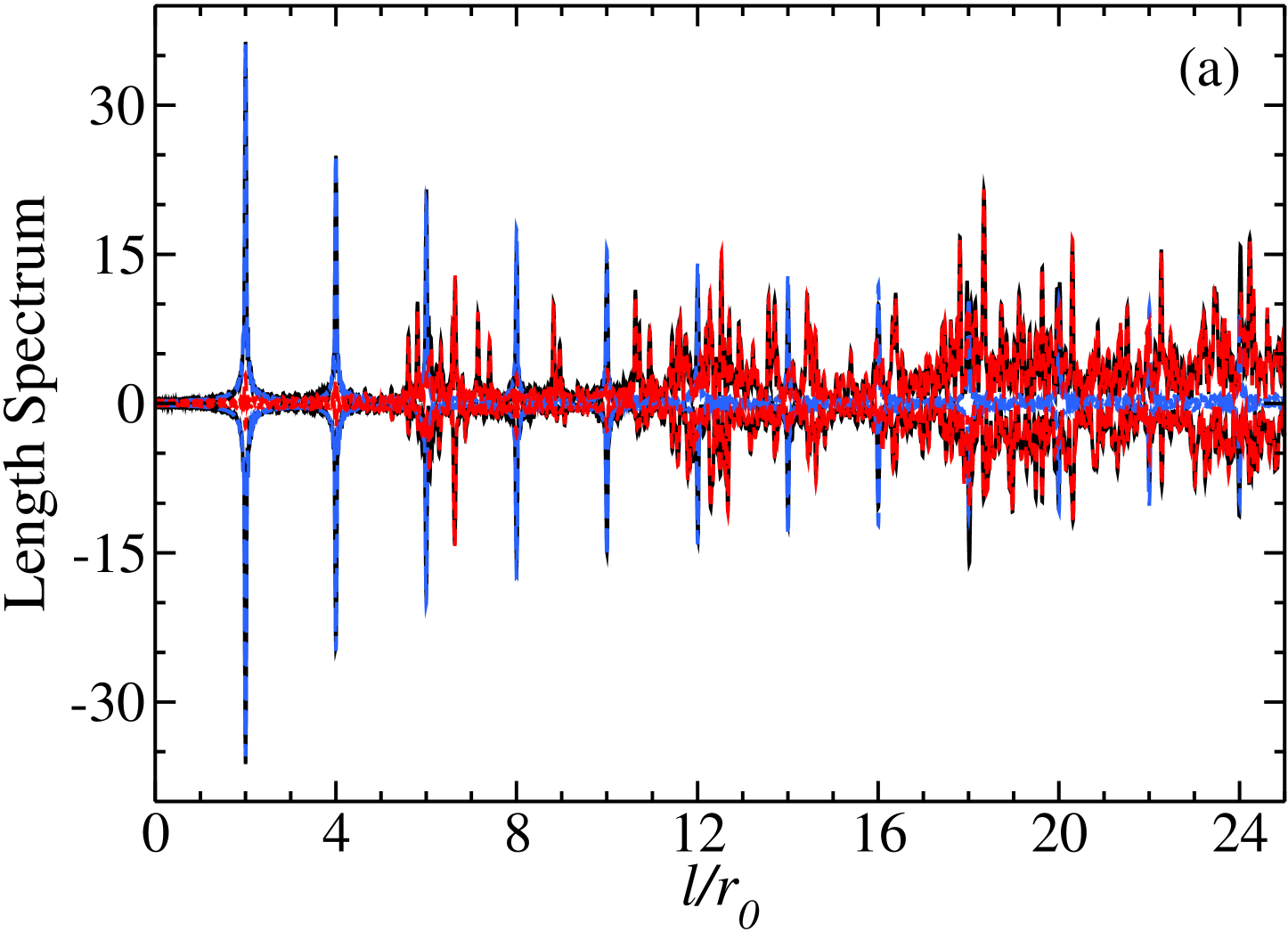}
	\includegraphics[width=\linewidth]{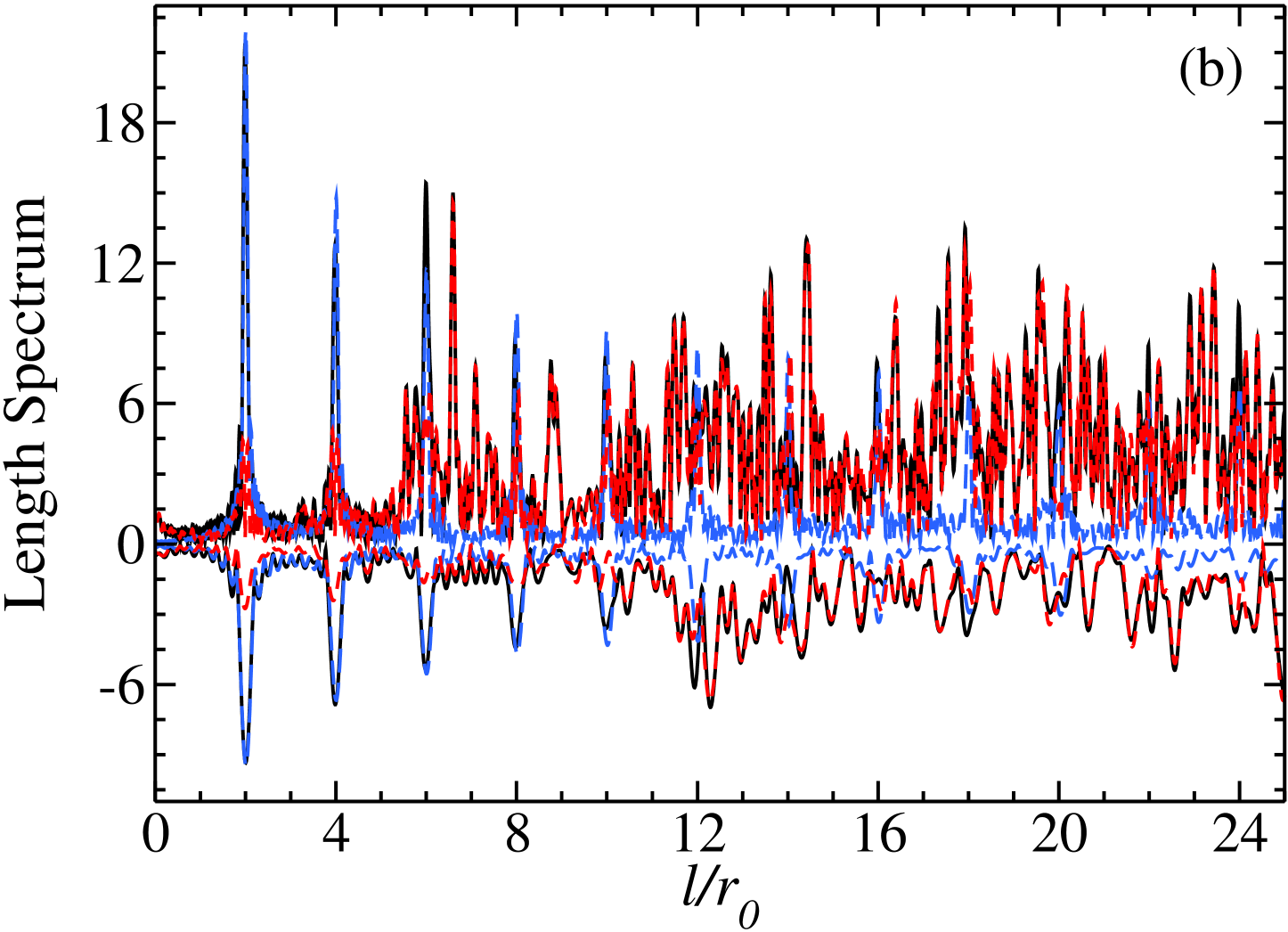}
		\caption{(a) Same as~\reffig{FFT_Full} (a) for the quarter stadium. (b) Same as~\reffig{FFT_Full} (b) for the quarter stadium GB.}\label{FFT_Quart}
	\end{figure}

We also analyzed the spectral properties. For this we unfolded the ordered wavenumbers $k_m$ to mean spacing unity by replacing them by the smooth part of the integrated spectral density, $\epsilon_m=N^{smooth}(k_m)$. For billiards it is given by Weyl's formula~\cite{Weyl1912}, $N^{smooth}(k_m)=\frac{\mathcal{A}}{4\pi}k_m^2+\gamma\frac{\mathcal{L}}{4\pi}k_m+C_0$, where we rescaled the wavenumbers such that the area equals $\mathcal{A}=4\pi$, for NRQBs $\gamma=-1$ and for massless NBs $\gamma=0$~\cite{Berry1987}. In~\reffig{Nfluc} the fluctuating part of the integrated spectral density, $N^{fluc}(k_m)=N(k_m)-N^{smooth}(k_m)$ is plotted. It exhibits fast fluctuations and slow oscillations. The latter originate from the BBOs. Indeed, in Ref.~\cite{Sieber1993} a procedure is developed for NRQBs which allows the computation of contributions of the BBOs and also the other periodic orbits shown in~\reffig{Scars} to the semiclassical trace formulas for the associated fluctuating part of the spectral density. It employs Gutzwiller's trace formula~\refeq{rhoNRQB} and was extended by one of the authors (B.D.) to NBs based on the trace formula~\refeq{rhoNB}~\cite{Zhang2021,Yupei2022,Dietz2022}.  The red curves show the analytical result for the BBOs~\cite{Sieber1993} for the NRQB
	\be
	N^{NRQB}_{bbo}(k)=\frac{a}{2\pi}\sqrt{\frac{k}{\pi R_0}}\sum_{m=1}^\infty m^{-3/2}\cos\left(2mkR_0-\frac{3\pi}{4}\right)
	\label{NbboNRQB}
	\ee
	and the NB
	\be
	N^{NB}_{bbo}(k)=\frac{a}{2\pi}\sqrt{\frac{k}{\pi R_0}}\sum_{m=1}^\infty m^{-3/2}\cos\left(2mkR_0 -m\pi -\frac{3\pi}{4}\right),
	\label{NbboNB}
	\ee
	respectively, with $R_0=r_0$ for the quarter stadium and $R_0=2r_0$ for the full stadium. They just differ by an additional phase of $m\pi$ and accordingly the quantization condition for the BBOs, which is obtained from the periodicity of the cosine functions, yields just a shift on the $k$ values of BBOs of the NB and NRQB with respect to each other. Generally, the quantization condition requires that $k$ multiplied with the length of the orbit on which the wave function is scarred (or its repititions) plus a possible phase, which depends on the BCs and for NBs also chirality, and can be read of the phases of the summands in the corresponding trace formula~\refeq{rhoNRQB} or~\refeq{rhoNB}, are multiples of $2\pi$. Thus, it is linear in the wave numbers, implying that the dispersion relation is not relevant. This can be seen when comparing~\reffig{Nfluc} (a) with (b) as outlined below in the sections on GBs and HGBs. In Figs.~\ref{FFT_Full} and~\ref{FFT_Quart} (a) we show the length spectra, that is, the Fourier transform from wavenumber to length of the fluctuating part of the spectral density $\rho^{fluc}(k_m)$ (black) of the full- and quarter-stadium NRQB and that of the NB multiplied with (-1), respectively. The turquoise curves show the Fourier transform of the fluctuating part of the spectral density originating from the BBOs deduced from Eqs.~\ref{NbboNRQB} and~\ref{NbboNB}, confirming their applicability even for finite values of $k_m$.  

\begin{figure}
	\centering
	\includegraphics[width=\linewidth]{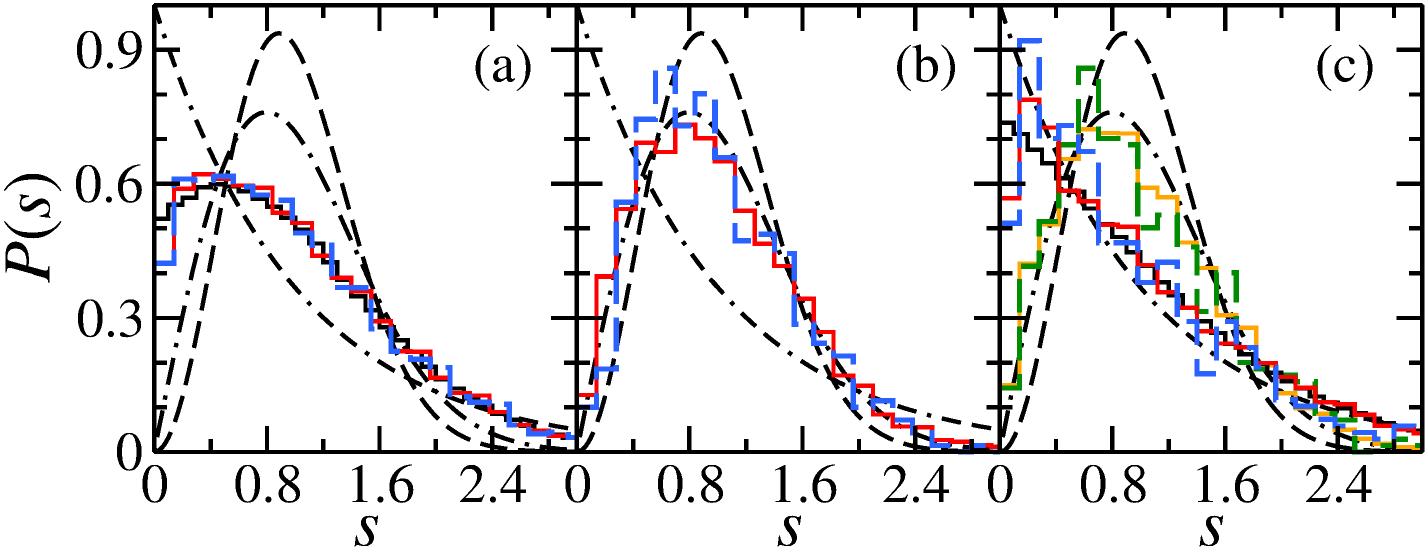}
		\caption{Nearest-neighbor spacing distributions for the full stadium. Shown are the distributions for (a) the NB before (blue histogram) and after (red histogram) extracting contributions from BBOs and for a random block-diagonal matrix with two blocks containing matrices drawn from the GOE (black histogram) that represent the two rotational-symmetry classes ($l=1$ and $l=1$), (b) for the symmetry-projected eigenstates ($l=0$) of the NB (red histogram) and for the HGB (blue histogram), (c) for the NRQB (red histogram) and the GB (blue histogram) and for a random block-diagonal matrix with four blocks containing matrices from the GOE (black histogram), that represent the four mirror-symmetry classes. In that panel, additionally the distributions of the quarter stadium NRQB (orange histogram) and GB (green histogram) are presented. In (b) and (c) only results after extraction of non-generic contributions are exhibited. The dash-dash-dotted lines, dash-dotted lines, dashed lines show the curves for Poisson, GOE and GUE statistics, respectively.   
	}\label{NND}
	\end{figure}

\begin{figure}
        \centering
        \includegraphics[width=\linewidth]{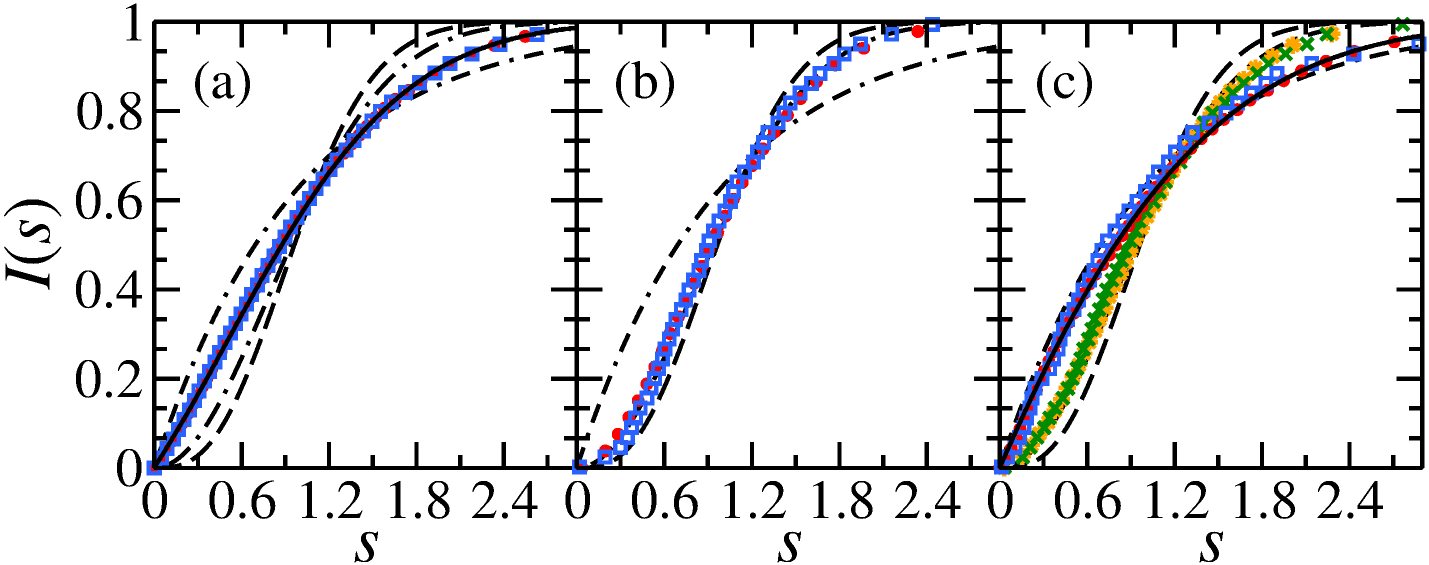}
		\caption{Same as~\reffig{NND} for the cumulants of the corresponding nearest-neighbor spacing distributions for the full stadium. Here, blue, red, orange and green histograms are replaced by blue squares, red dots, orange stars and green crosses, respectively. 
        }\label{INND}
        \end{figure}

\begin{figure}
	\centering
	\includegraphics[width=\linewidth]{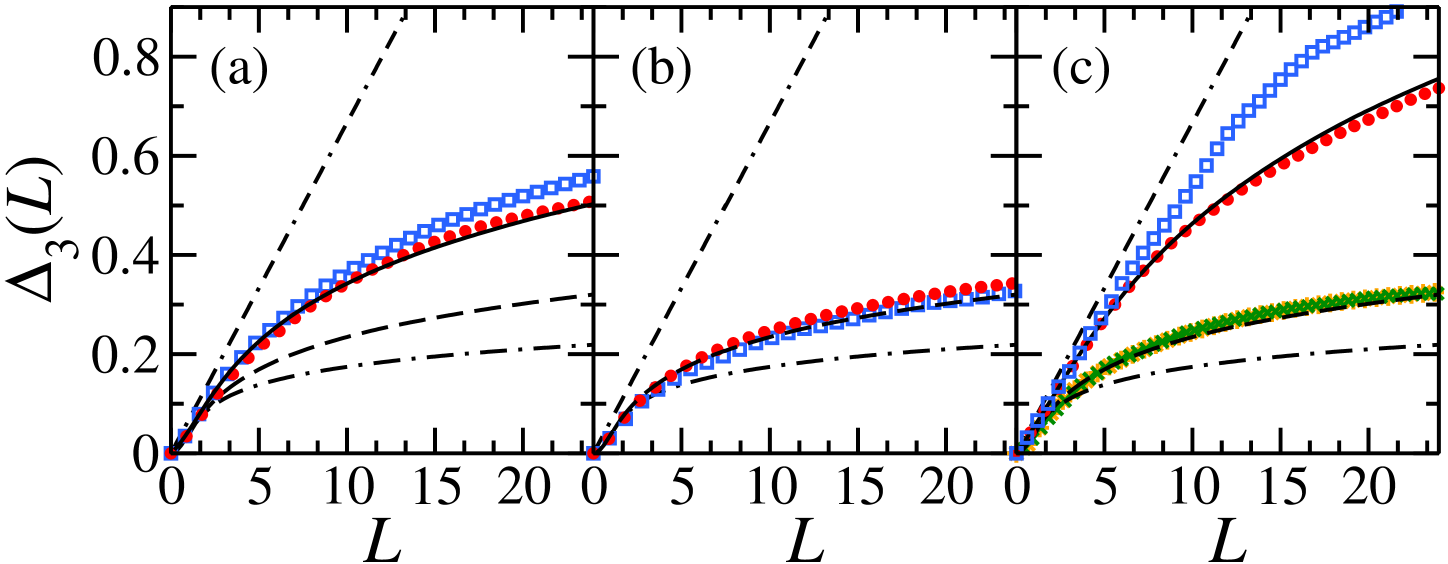}
	\caption{Same as~\reffig{INND} for the spectral rigidity $\Delta_3(L)$.
	}\label{Delta3}
	\end{figure}

	In Figs.~\ref{NND}-\ref{Delta3} (a) are shown the nearest-neighbor spacing distribution $P(s)$, its cumulant $I(s)=\int_0^sd\tilde sP(\tilde s)$ and the spectral rigidity $\Delta_3(L)$~\cite{Bohigas1975,Mehta2004} for the complete eigenvalue spectrum of the full-stadium NB (blue histograms and squares), obtained by uniting those for the symmetry classes $l=0$ and $l=1$. We find that $P(s)$ and I(s) agree well with that of block-diagonal random matrices consisting of two blocks formed by matrices from the GOE, each one representing one symmetry class, whereas for the $\Delta_3$ statistics clear deviations are observed. These have been attributed to BBOs for the NRQB~\cite{Graef1992}, where the eigenvalue spectrum of a quarter stadium was determined experimentally with a microwave billiard of that shape. Actually, the effect of BBOs on the spectral properties was detected in these experiments and the foundations for their semiclassical treatment and extraction from eigenvalue spectra were layed in Ref.~\cite{Sieber1993} where, actually, trace formulas have been derived for all orbit families shown in~\reffig{Scars}. These can be employed to extract non-generic contributions from the spectrum, by taking into account the slow oscillations caused by them. This is achieved by unfolding with the sum of the smooth part, i.e. the Weyl term and the slow oscillations which are described by these trace formulas, $\epsilon_m=N^{smooth}(k_m)+N_{BBO}(k_m)+...$. Like for NRQBs the agreement with the random-matrix predictions is already very good, when taking into account only contributions from the BBOs ; cf. red histogram and dots in Figs.~\ref{NND}-\ref{Delta3} (a). We, in addition, computed the trace formulas provided in Ref.~\cite{Sieber1993} for the other orbits in~\reffig{Scars}, and confirmed that also for the NB their contributions are negligible, as found in that work for NRQBs. This is to be expected, because according to the momentum distributions and Husimi functions localization is strongest for these orbits and almost BBOs. Furthermore, to confirm that contributions from the BBOs are completely extracted we compare in Figs.~\ref{FFT_Full} and~\ref{FFT_Quart} the length spectra obtained from the Fourier transform of $\tilde\rho^{fluc}(k_m)=\rho^{fluc}(k_m)-\rho^{XB}_{bbo}(k_m)$ (red dashed lines) to the length spectrum deduced from $\rho^{XB}(k_m)$ (black). Here, $XB=NRQB,NB$. Indeed, peaks located at the lengths of the BBOs are absent or suppressed in the former one.  

	\section{Graphene and Haldane Billiards\label{GBHB}}
	\subsection{Brief review of the salient differences of GBs and HGBs}
	The honeycomb lattice of graphene is formed by two interpenetrating triangular sublattices $A$ and $B$; cf. left part of~\reffig{BandStr_GB}. The basis vectors $\mathbf{a}_i$ are defined as
	$
	\mathbf{a}_1=(1,0),  \mathbf{a}_2=(-\frac{1}{2}, \frac{\sqrt{3}}{2}),  \mathbf{a}_3=(-\frac{1}{2}, -\frac{\sqrt{3}}{2}) 
	$,
	where the distance $a$ between neighboring sites of the honeycomb lattice is set equal to unity, $a=1$. In quasi-momentum space, the Hamiltonian is given by 
	\begin{equation}
		\hat H^0(\boldsymbol{\kappa}) = t_1 {\sum}_{i=1}^3  \big( \hat\sigma_x \cos( \boldsymbol{\kappa} \cdot \mathbf{a}_i)  - \hat\sigma_y \sin(\boldsymbol{\kappa} \cdot \mathbf{a}_i) \big) ,
	\end{equation}
	in the basis $\psi^T=\left(\psi_A(\boldsymbol{\kappa}),\psi_B(\boldsymbol{\kappa})\right)$~\cite{Dresselhaus1996} where $\hat\sigma_x,\hat\sigma_y,\hat\sigma_z$ denote the Pauli matrices. The conduction and valence band~\cite{Neto2009} are shown in the right part of~\reffig{BandStr_HB}.
	Near the Dirac cone $\mathbf{K}=\left(\frac{2\pi}{3 a},\frac{2\pi}{3 \sqrt{3}a}\right)$ the effective Hamiltonian is given by the Dirac Hamiltonian for a spin-1/2 particle,
		\begin{equation}
		   \hat H^0_K(\mathbf{q})=\frac{3 t_1 a}{2}\boldsymbol{\hat\sigma}\cdot\mathbf{q}, \qquad (\boldsymbol{\kappa}= \mathbf{K}+ \mathbf{q})
		\end{equation}
	with $\boldsymbol{\hat\sigma}=({\hat\sigma}_x,{\hat\sigma}_y)$ and $\boldsymbol{q}$ the quasimomentum vector.
	Similarly, near the Dirac cone $\mathbf{K}'=-\mathbf{K}$ the effective Hamiltonian becomes
	\begin{equation}
	\hat H^0_{K'}(\mathbf{q})=\frac{3 t_1 a}{2} \boldsymbol{\hat\sigma^\ast}\cdot\mathbf{q}, \qquad (\boldsymbol{\kappa}= \mathbf{K}'+ \mathbf{q}),
	\end{equation}
	where $\boldsymbol{\hat\sigma^\ast}$ denotes complex conjugation of $\boldsymbol{\hat\sigma}$. The GBs are constructed by cutting their shape out of a honeycomb-lattice sheet and imposing the Dirichlet BC along the outer boundary.  

	The extension to the Haldane model is achieved by adding a nonzero purely imaginary next-to-nearest neighbor tunneling parameter, $i t_2$~\cite{Haldane1988}, as illustrated in the left part of~\reffig{BandStr_HB}. Here, the next-to-nearest neighbor site vectors are $\mathbf{b}_1=(0,\sqrt{3} a), \mathbf{b}_2=(-\frac{3 a}{2}, -\frac{\sqrt{3}a}{2}),\mathbf{b}_3=(\frac{3 a}{2}, -\frac{\sqrt{3}a}{2})$. Even though complex tunneling explicitly breaks time reversal invariance, the total magnetic flux through a unit cell is zero. Furthermore, we introduce onsite potentials $M$ with $M>0$ on all sites of sublattice $A$ and $-M$ on all sites of sublattice $B$, yielding in the quasi-momentum space the Hamiltonian 
	\begin{equation}
	\label{eq:HamHD}
	\hat H(\boldsymbol{\kappa})=\hat H^0(\boldsymbol{\kappa})+ \left(M- 2 t_2 \sum_{j=1}^3 \sin (\boldsymbol{\kappa}\cdot \mathbf{b}_j)  \right)\hat\sigma_z,
	\end{equation}
	where the first and second term in the rounded brackets induce breaking of the inversion symmetry and violation of time-reversal invariance, respectively. Near the $K$ and the $K'$ points, respectively, the effective Hamiltonian equals 
		\begin{align}
			\hat H_K(\mathbf{q})  &=\frac{3 t_1 a}{2}\boldsymbol{\hat\sigma}\cdot\mathbf{q} + \left(M- 3 \sqrt{3}t_2\right)\hat\sigma_z \\
			\hat H_{K'}(\mathbf{q})  &=\frac{3 t_1 a}{2}\boldsymbol{\hat\sigma^\ast}\cdot\mathbf{q} + \left(M+ 3 \sqrt{3}t_2\right)\hat\sigma_z .
		\end{align}
	Both Dirac cones are gapped in the insulator phase, $|t_2| < \frac{M}{3\sqrt{3}}$, and the Chern insulator-phase $|t_2| > \frac{M}{3\sqrt{3}}$ defining a non-trivial topological phase with non-zero Chern number. On the contrary at the critical point $t_2=\frac{M}{3\sqrt{3}}$ only one Dirac cone, namely that at the $K'$ point is gapped with the effective mass $2M$ and that at the $K$ point survives, implying that in the low energy limit $|E|< M$, there is only one Dirac cone. Similarly for $t_2=-\frac{M}{3\sqrt{3}}$, the Dirac cone at the $K$ point is gapped out and a single Dirac cone remains at the $K'$ point. We have demonstrated in Ref.~\cite{Nguyen2024} that at these critical points the fluctuation properties in the energy spectrum of a Haldane graphene billiard (HGB), which is obtained by applying the Haldane tunneling parameter and mass terms to a GB of corresponding shape, coincide in the energy window $|E|\leq{\rm min}(\vert t_1\vert/2,M)$ with those of the corresponding NB. 

	\subsection{Properties of the eigenstates of the stadium GB and HGB\label{SpectrGBHB}}
	We constructed quarter- and full-stadium GBs and HGBs by cutting their shape out of a honeycomb lattice, which was oriented such the straight parts coincide either with an armchair edge or a zigzag edge. We present only results for the latter case. In the numerical simulations we used the Haldane model at the critical point $t_2=\frac{M}{3\sqrt{3}}$ and set $t_1=1$  and $M=0.3$, so that ${\rm min}(\vert t_1\vert/2,M)=0.3$. Furthermore, for the computation of the eigenvalues of the stadium GB and HGB with the same geometry $\frac{L}{r_0}=\frac{1+\sqrt{5}}{2}$ as for the NRQB and NB, we used $N\approx 130000$ sites, and for the computation of wave-functions $N\approx 90000$. To determine the eigenstates we proceeded as in Ref.~\cite{Dietz2015} and diagonalized the $N\times N$-dimensional tight-binding Hamilton matrix in configuration space, yielding the ordered-by-size eigenvalues and wave function components at the lattice sites as the corresponding eigenvector components. 

\begin{figure}
	\centering
	\includegraphics[width=\linewidth]{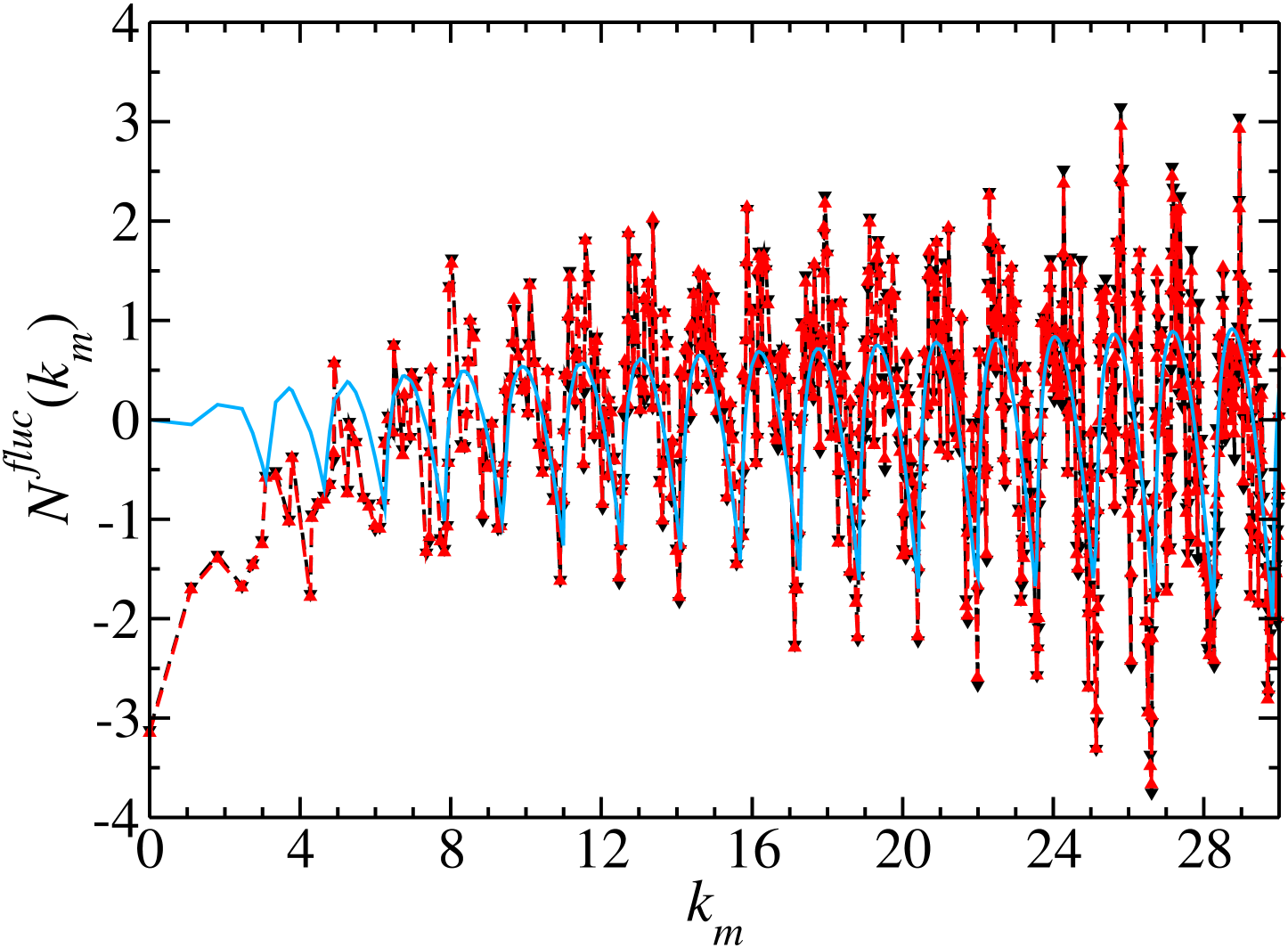}
	\includegraphics[width=\linewidth]{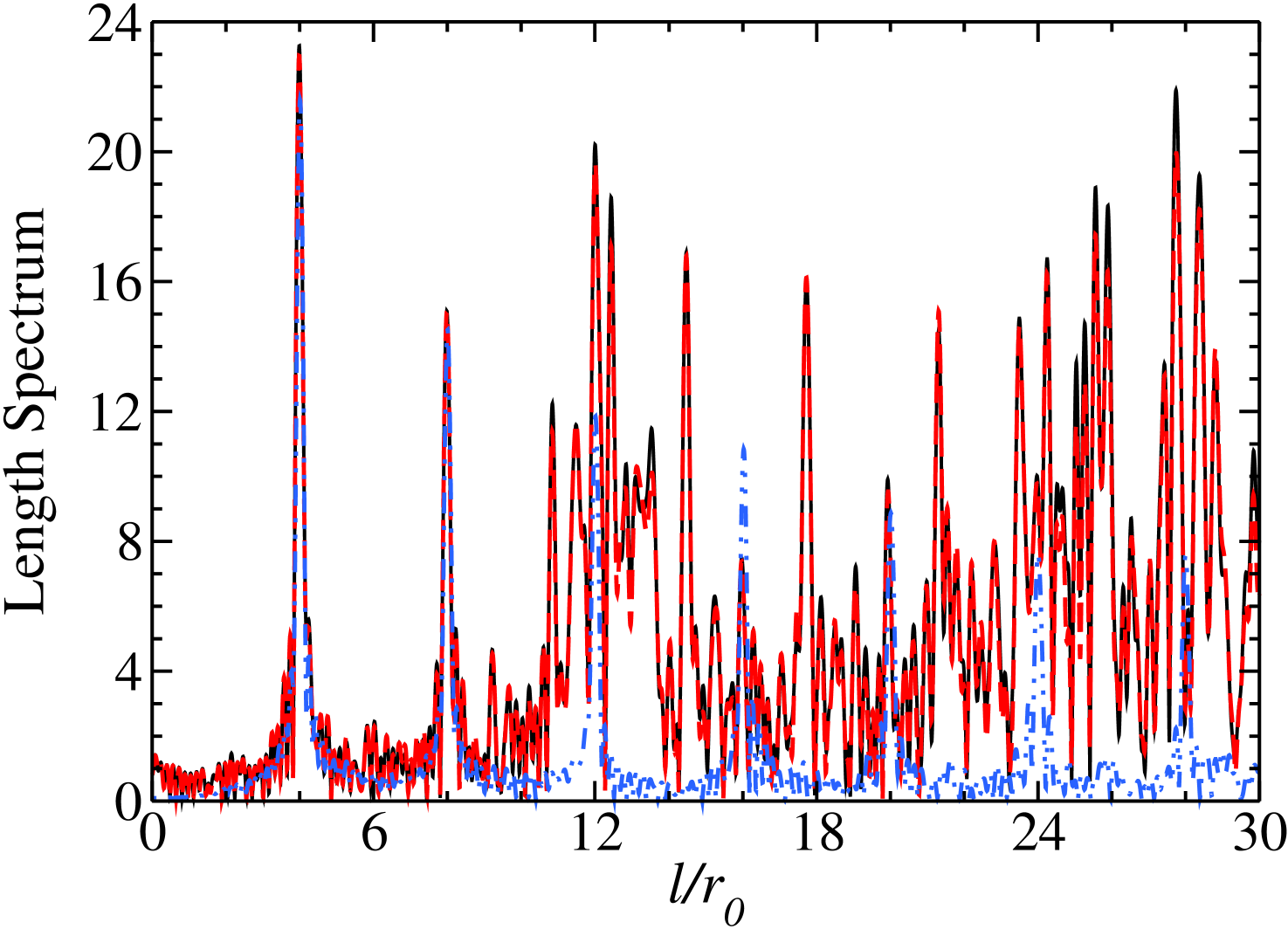}
		\caption{(a) Fluctuating part of the integrated spectral density of the full stadium GB (black) and HGB (red) at the lower band edge. The blue curve is obtained from the trace formula for the BBOs~\refeq{NbboNRQB} of the NRQB. (b) Same as (a) for the length spectra.}\label{Nfluc_FFT_GBHB}
	\end{figure}

	We analyzed spectral properties starting from the lower band edge and the Dirac point, respectively. For the unfolding, we used the same procedure as in Ref.~\cite{Zhang2021}, namely we shifted the eigenvalues $E_i$ such that at the band edge or at the Dirac point $\tilde E_1 = 0$, that is, $\tilde E_m=E_m-E_1$. Then we replaced them by the smooth part of the integrated spectral density, $\epsilon_m=N^{smooth}(\tilde k_m)$ with $\tilde k_m$ denoting the effective wavenumbers, where $\tilde k_m=\sqrt{\tilde E_m}$ at the band edge and $\tilde k_m=\tilde E_m$ at the Dirac point. To ensure that the correspondance to the stadium NRQB is preserved around the band edges and that only wavenumbers within the region of linear dispersion are considered, we fit a polynomial of second order in $k$, $N^{smooth}(k)=a_0+a_1k+a_2k^2$ to the integrated spectral density. We thereby and through additional tests found out that even for the larger lattice structure only $\approx 1200$ and $\approx 500$ eigenstates around the band edge and Dirac point, respectively, comply with these requirements. For the GB there are $\approx 100$ nearly-degenerate edge states above the Dirac point for the GB and just one for the HGB, which we disregarded. Accordingly, for the GB the lowest $\approx 20$ eigenvalues are missing. For a direct comparison with properties of the eigenstates of the NRQB and NB we rescaled the effective wavenumber, $k_m=\sqrt{a_2}\tilde k_m$ such that the scale of $k_m$ is the same for all billiards, as can be seen in~\reffig{Nfluc_FFT_GBHB} and~\reffig{Nfluc}. 

	In addition to considering the full-stadium GB and NRQB we also analyzed spectral properties of the fully disymmetrized full stadium billiard, that is the quarter-stadium GB and NRQB. The reason is that, while for the NRQB we can obtain the eigenstates for the two symmetry classes separately, this is not possible in the GB, because the mirror symmetry is perturbed along the axis with zigzag-edge structure, implying that a complete extraction of contributions from BBOs is not possible in that case; cf. Figs.~\ref{NND}-\ref{Delta3} (c). In~\reffig{Nfluc_FFT_GBHB} the fluctuating part of the integrated spectral density and the length spectra of the full stadium GB (black) and HGB (red) are compared around the band edges. They lie on top of each other, also for the quarter-stadium GB and HGB (not shown). Like in the NRQB and NB we observe slow oscillations which are indeed well described by the semiclassical trace formula~\refeq{NbboNRQB} of the NRQB (blue curves). We would like to mention, that at the band edges $\mathcal{N}^{fluc}(k_m)$ and the length spectra also agree very well with those of the corresponding NRQB, thus confirming the findings of Ref.~\cite{Dietz2015}. Therefore, we consider in the following only the properties of the eigenstates around the Dirac point. 

	In~\reffig{Nfluc} (a) we compare the fluctuating part of the integrated spectral density of the quarter-stadium GB around the Dirac point (blue squares) with that of the corresponding NRQB and the semiclassical trace formula~\refeq{NbboNRQB} for its BBOs. Similarly, in~\reffig{Nfluc} (b) is shown that of the quarter-stadium HGB together with that of the NB and the semiclassical result for its BBOs,~\refeq{NbboNB}. The slow oscillations observed for the GB are well described by~\refeq{NbboNRQB} and those of the HGB by~\refeq{NbboNB}. This gives a first hint, that the scarred states observed in the GB exhibit nonrelativistic features, and those of the HGB behave like relativistic ones. In Figs.~\ref{NND}-\ref{Delta3} (b) and (c) are shown the nearest-neighbor spacing distribution, its cumulant and the spectral rigidity of the full stadium HGB and GB (blue histograms and squares), respectively. For the HGB they agree well with those of the NB after separation into the symmetry classes $l=0$ and $l=1$ for which spectral properties are the same. Shown are the curves for the case $l=0$ (red histograms and dots). This good agreement reflects the fact that the addition of Haldane tunneling and onsite-potential of opposite sign on the sublattices $A$ and $B$ of a hexagonal lattice leads to a breaking of the twofold symmetry~\cite{Nguyen2024,Zhang2021,Zhang2023c}. That is, like the eigenspinors of the full-stadium NB, the eigenstates of the HGB cannot be classified according to their transformation properties under a rotation by $\pi$. On the contrary, for the GB we find good agreement with the spectral properties of the full-stadium NRQB, that is, with those of random block-diagonal matrices consisting of four GOE blocks, implying that its spectrum consists of four spectra associated with the mirror-symmetries with respect to the $x$ (zigzag) and $y$ (armchair) edges. This again is a feature of the NRQB but not conform with the symmetry properties of the solutions of the Dirac equation~\refeq{DE} and thus of NBs as outlined above in the section on NRQBs and NBs. Due to the slight perturbation of the mirror-symmetry with respect to the $x$ axis, degeneracies are lifted, so that a complete extraction of non-generic contributions from scarred eigenstates is not possible. Therefore, deviations are observed in the long-range correlations, however, the nearest-neighbor spacing distribution and its cumulant agree well with that of the full-stadium NRQB. To corroborate this reasoning, we in addition show in Figs.~\ref{NND}-\ref{Delta3} (c) the spectral properties of the quarter-stadium NRQB (orange histogram and diamonds) and GB (green histogram and circles) after extraction of non-generic contributions and find very good agreement. 

	Furthermore, in Figs.~\ref{FFT_Full} and~\ref{FFT_Quart} (b) we compare length spectra of the full-stadium HGB and of the quarter-stadium GB (black) with those obtained from Eqs.~\ref{NbboNB} and~\ref{NbboNRQB}, respectively, and with the length spectra of the corresponding $\tilde\rho^{fluc}(k_m)=\rho^{fluc}(k_m)-\rho^{fluc}_{BBO}(k_m)$, which leads to a considerable suppression of the peaks at lengths of BBOs. The upper part shows the length spectrum deduced from the eigenvalues around the band edges in the nonrelativistic regime, the lower part that obtained in the region of linear dispersion around the Dirac point. Comparison of these spectra reveals, that both at the band edges and at the Dirac point the length spectra exhibit peaks associated with BBOs of comparable relative heights as in the NRQB and NB, however, generally the peak structure differs especially for the quarter-stadium GB around the Dirac point, in the sense that additional peaks appear. This is attributed to the edge structure of the lattice.  

\begin{figure}
	\centering
	\begin{tabular}{@{}c@{}}
	\begin{tabular}{@{}c@{}}
	\includegraphics[width=.42\linewidth]{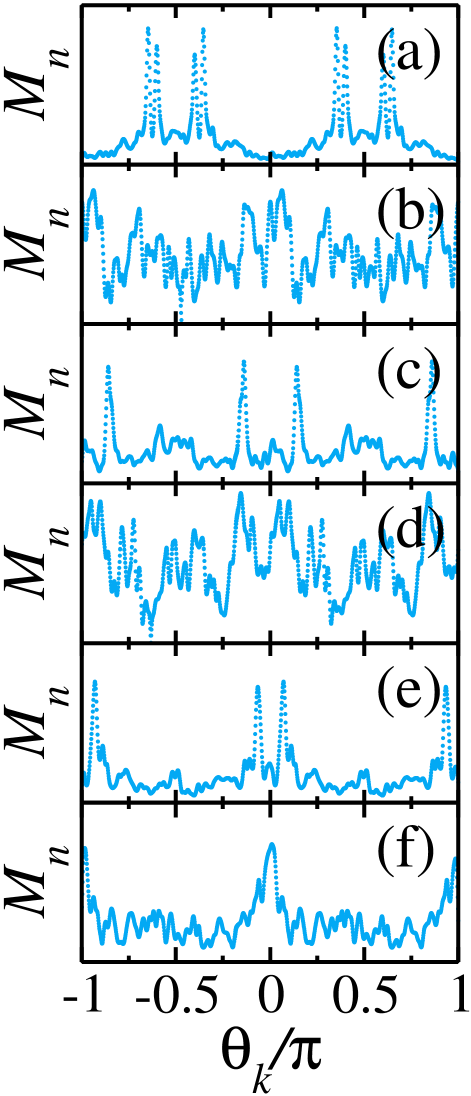}
	\end{tabular}
	\begin{tabular}{@{}c@{}}
	\includegraphics[width=.355\linewidth]{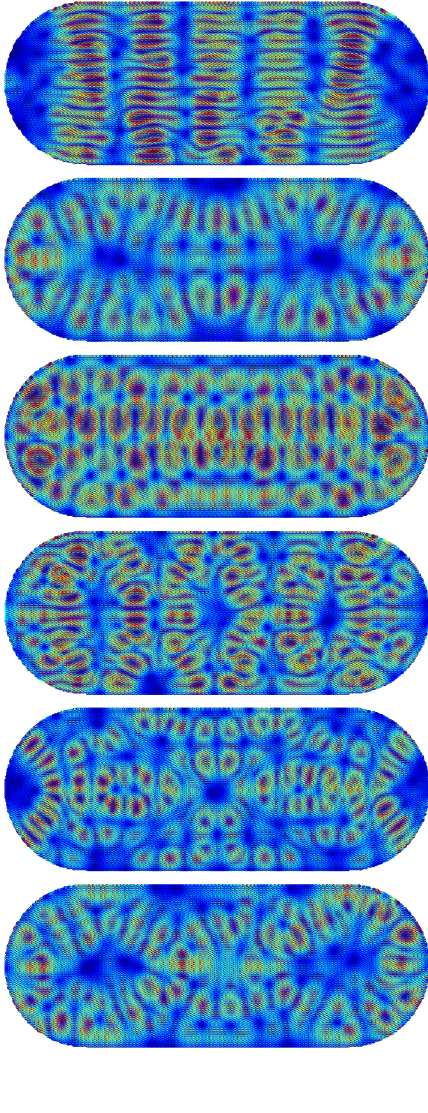} \\[\abovecaptionskip]
	\end{tabular}
	\end{tabular}
		\caption{On-shell momentum distribution of eigenstates of the GB above the Dirac point versus momentum direction $\theta_k$ for (a) $m=274$,  (b) $m=117$, (c) $m=242$, (d) $m=122$, (e) $m=220$, (f) $m=182$. To the right are exhibited the corresponding wave functions. Note, that statenumbers $m$ correspond to similar wavenumbers $k_m$ in the NRQB and GB.}\label{Mom_GB}
	\end{figure}

\begin{figure}
	\centering
	\begin{tabular}{@{}c@{}}
	\begin{tabular}{@{}c@{}}
	\includegraphics[width=.42\linewidth]{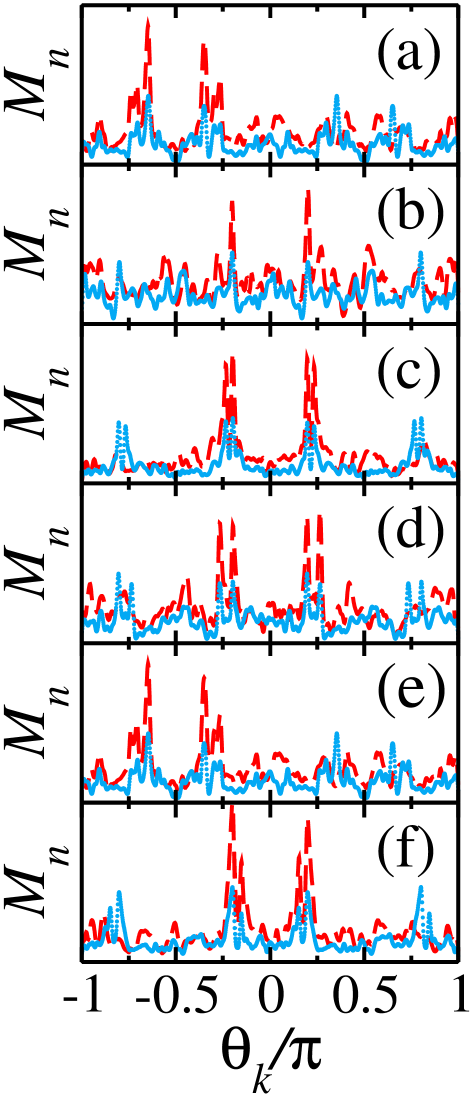}
	\end{tabular}
	\begin{tabular}{@{}c@{}}
	\includegraphics[width=.355\linewidth]{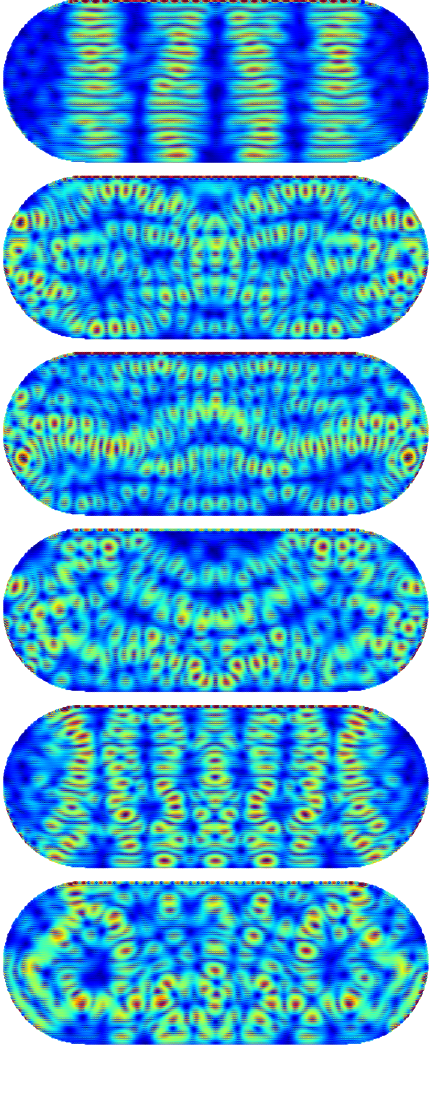} \\[\abovecaptionskip]
	\end{tabular}
	\end{tabular}
		\caption{On-shell momentum distribution of eigenstates of the HGB above the Dirac point versus momentum direction $\theta_k$ for (a) $m=201$, (b) $m=278$, (c) $m=257$, (d) $m=264$, (e) $m=223$, (f) $m=217$. To the right are exhibited the corresponding wave functions. Note, that statenumbers $m$ correspond to similar wavenumbers $k_m$ in the NB and HGB.}\label{Mom_HB}
	\end{figure}

\begin{figure}
	\centering
	\includegraphics[width=.9\linewidth]{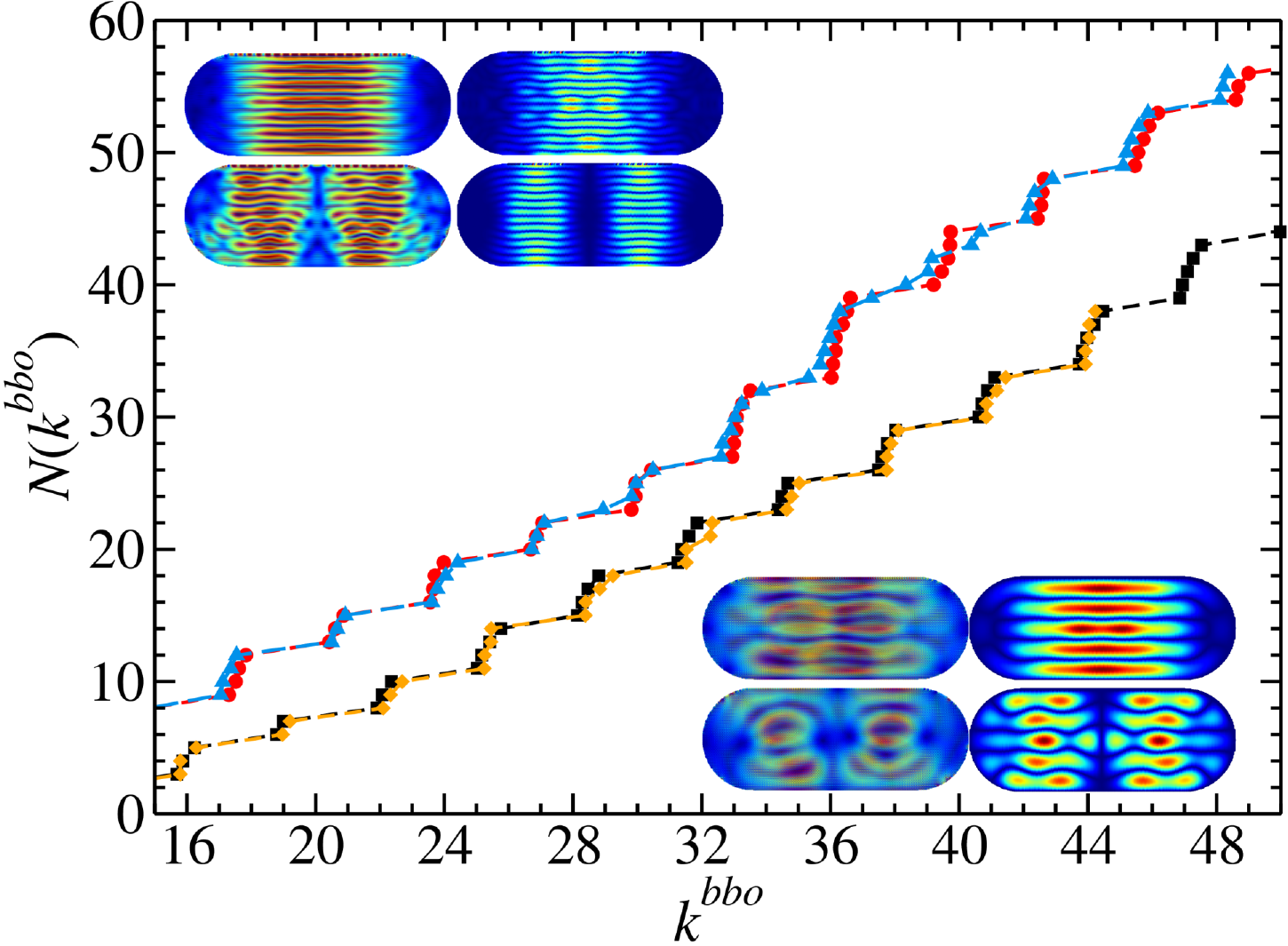}
\caption{Number of BBOs of the full-stadium NRQB (black) and NB (red) versus the corresponding wavenumber compared to those of the GB (orange) and HGB (blue). The upper inset shows wave functions of the HGB and local current of the NB for the two lower values of $k^{bbo}$ of the triplet around $k^{bbo}\simeq 20$, the lower one for the wave functions of the GB and QB for the upper values of the first triplet around $k^{bbo}\simeq 16$.}\label{Num_BBOs}
\end{figure}

	In Figs.~\ref{Mom_GB} and~\ref{Mom_HB} are shown examples of wave functions of the GB and HGB, respectively, with similar scarring structure as in those shown in Figs.~\ref{Mom_QB} and~\ref{Mom_NB}. Note, that the wave functions of the GB are real, not complex as is the case for the solutions for the Dirac equation~\refeq{DE} and exhibit a mirror symmetry with respect to the $y$ (armchair) edge and an almost mirror-symmetry along the $x$ (zigzag) axis. Those of the HGB are complex, and only their modulus exhibits an almost mirror-reflection symmetry with respect to the $y$ axis, whereas there is no symmetry with respect to the $x$ axis. Similar to the momentum distributions of the NRQB and NB shown in Figs.~\ref{Mom_QB} and~\ref{Mom_NB} for both the GB and HGB the momentum dsistributions exhibit a few sharp peaks for BBOs around $\theta_k/\pi=\pm 1/2$ shown in Figs.~\ref{Mom_GB} (a) and~\ref{Mom_HB} (a), (e) corresponding to slightly different momentum directions in the stripes. For wave functions localized around orbits reflected at the semicircular parts close to the center as in ~\reffig{Mom_GB} (b), (c) and (f) they are also localized around zero for the GB, but in~\reffig{Mom_GB} (c) and for the HGB in~\reffig{Mom_HB} (b)-(d) they are more pronounced along other clearly discernible wave-patterns localized on trajectories that are reflected at the semicricular part. In Figs.~\ref{Mom_GB} and~\ref{Mom_HB} (e) they are similar to the bow-tie like orbits or almost BBOs like in~\reffig{Mom_QB}) and~\ref{Mom_NB} (e), respectively. Yet, in all examples the wave function patterns and the peak structure of the momentum distributions are not as clearly discernible as in Figs.~\ref{Mom_QB} and~\ref{Mom_NB}, the reason being that the attainable wave numbers are not high enough. Analysis of the participation numbers~\refeq{IPR} yields largest values for the BBOs, as in the case of the NB and NRQB; cf.~\reffig{IPR_Mom_QB_NB}. Generally, for the HGB even the modulus of the wave functions does not exhibit the mirror symmetry with respect to the $x$ axis observed for the NB in~\reffig{Mom_NB}. This manifets intself in the momentum distributions when using the complex wave functions (red curves in~\reffig{Mom_HB}) instead of just the real part (blue curves in~\reffig{Mom_HB}). Indeed, as indicated by the spectral properties, for the HGB the combination of the Haldane tunneling term and the mass term, which differs by a sign for the $A$ and $B$ lattices, applied to the hexagonal structure destroys the two-fold symmetry resulting in an asymmetric combination of the two symmetry classes. We also analyzed Husimi functions, however the values of $k$ are to low to obain any useful tool for the identification of quantum-scarred wave functions. Since only $\approx 500$ eigenstates are available, it is not needed. Still, the analogy to the NB remains as clearly visible in~\reffig{Num_BBOs}, namely the $k$ values for the BBOs of the GB and HGB agree very well with those of the NRQB and NB, respectively, except for a few values for the latter case. The upper and lower insets compare two examples of wave functions for the HGB and NB, and the GB and QB, respecively.

	\begin{figure}
	\centering
	\includegraphics[width=.9\linewidth]{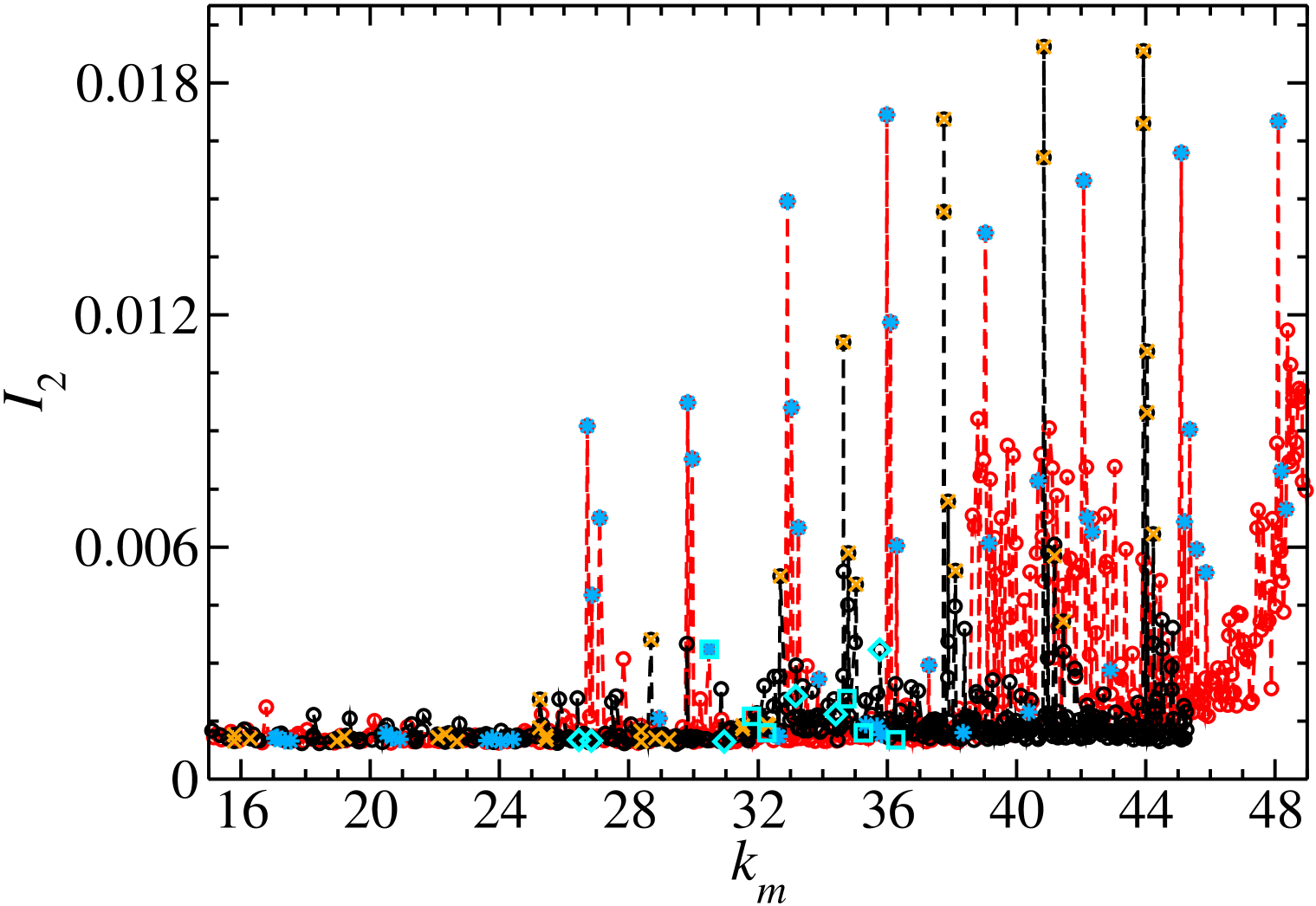}
		\caption{Participation numbers of the on-shell momentum distributions of eigenstates of the GB (black) and HGB (red). Orange crosses show the participation numbers for the BBOs of the GB, blue stars those of the HGB. Examples for such wave functions are shown in Figs.~\ref{Mom_GB} (a) and ~\ref{Mom_HB} (a), respectively. The participation numbers of the momentum distributions shown for the GB in~\reffig{Mom_GB} and the HGB in~\reffig{Mom_HB} are marked by cyan diamonds and squares, respectively.
	}\label{IPR_GBHB}
	\end{figure}

\section{Conclusions and Discussion\label{Concl}}
We present a thorough study of properties of quantum-scarred eigenstates of nonrelativistic quantum billiards, relativistic neutrino billiards, and of graphene and Haldane-graphene billiards and their impact on the spectral properties. Here, for the GB and HGB the region of linear dispersion around the Dirac point is of main interest. The billiards have the shapes of a full and a quarter stadium. We demonstrate that the scarred eigenstates of the GB do not exhibit the features of a relativistic quantum billiard, but agree with those of the NRQB, whereas for the HGB we find good agreement with those of the NB. Here, we employ RMT and the semiclassical approach for the analysis of the spectral properties, and momentum-distributions and for the NRQB and NB also Husimi functions to identify and classify scarred eigenstates. We find that the nongeneric contributions to the spectral density of the GB are well described by the semiclassical trace formula for the corresponding NRQB derived in Ref.~\cite{Sieber1993}, whereas that of the HGB and NB are well captured by the semiclassical approach for relativistic NBs developed in Ref.~\cite{Dietz2020,Dietz2022}. Furthermore, the spectral properties of the full-stadium GB agree well with those of the NRQB, of which the spectrum is a superposition of the spectra associated with the four mirror-symmetry classes and is well described by block-diagonal random matrices, consisting of four blocks of matrices from the GOE. For the quarter stadium billiard we find for the GB and NRQB good agreement with GOE statistcs and the the NB and HGB with GUE statistics. This indicates that the GB like the NRQB has mirror symmetries with respect to the $x$ and $y$ axes. Indeed, the wave functions of the GB exhibit a mirror-symmetry along the central armchair axis and a slightly perturbed mirror symmetry along the central zigzag axis, and they are real for both billiard systems. On the other hand, the fluctuating part of the spectral density resulting from scarred eigenstates of the HGB agrees very well with that of NBs of corresponding shape, the wave functions are complex and don't exhibit mirror symmetries as expected for the solutions of the Dirac equation~\refeq{DE}. We also demonstrate that for all considered billiard systems only eigenstates that are scarred along BBOs have a perceivable impact on the spectral properties. We checked this by employing the semiclassical trace formulas provided in Ref~\cite{Sieber1993} for the orbits on which the eigenstates are scarred (cf.~\reffig{Scars}) for NRQBs and derived by following the methods of Ref.~\cite{Sieber1993} based on Refs.~\cite{Dietz2020,Dietz2022} for NBs. We come to the conclusion that contrary to the general belief~\cite{Huang2009,Ge2024} the quantum scars observed in GBs -- as demonstrated for the full- and quarter-stadium graphene billiards -- do not exhibit the features, like chirality, of relativistic QBs~\cite{Zhang2021,Zhang2023c} even though they are effectively governed by the Dirac equation of relativistic spin-1/2 particles. On the contrary, HGBs exhibit the features required for relativistic quantum scars. Consequently, even though GBs exhibit relativistic phenomena, they do not provide a suitable model for the study of aspects of relativistic quantum chaos. 

The recent proposal to simulate the Haldane model with photonic crystals~\cite{Jotzu2014,Liu2021,Lannebere2019} provides a possibibility to realize energy spectra exhibiting the particular features of NBs and thus of relativistic quantum scars experimentally. Yet, although the Haldane model has been realized experimentally on several platforms, observing quantum chaos in these systems remains challenging due to limitations in system size, boundary control, and the precise measurement of an energy spectrum. Nevertheless, exploring alternative platforms for realizing the Haldane model represents a fruitful direction for future research. 

\section{Acknowledgement} 
We acknowledge financial support from the Institute for Basic Science in Korea through the project IBS-R024-D1.
T. \v{C}. was supported by an appointment to the JRG Program at the APCTP through the Science and Technology Promotion Fund and Lottery Fund of the Korean Government and by the Korean Local Governments - Gyeongsangbuk-do Province and Pohang City.
\bibliography{References}
\end{document}